\documentstyle[12pt,epsfig]{article}
\newcommand{\blankline}{\vskip .3cm}
\newcommand{\f}{\begin{equation}}
\newcommand{\ff}{\end{equation}}
\newcommand{\be}{\begin{equation}}
\newcommand{\ee}{\end{equation}}
\newcommand{\bea}{\begin{eqnarray}}
\newcommand{\eea}{\end{eqnarray}}
\begin{document}
\centerline{\LARGE New varying speed of light theories}
\blankline \blankline \rm \centerline{Jo\~ao Magueijo}\blankline
\centerline{\it The Blackett Laboratory,Imperial College of
Science, Technology and Medicine } \centerline{\it South
Kensington, London SW7 2BZ, UK}

\blankline
\blankline
\blankline
\blankline
\centerline{}
\blankline
\blankline
\blankline
\blankline

\centerline{ABSTRACT}
\blankline
We review recent work on the possibility of
a varying speed of light (VSL). We start by discussing the
physical meaning of a varying $c$, dispelling the myth that the
constancy of $c$ is a matter of logical consistency. We then
summarize the main VSL mechanisms proposed so far: hard breaking
of Lorentz invariance; bimetric theories (where the speeds of
gravity and light are not the same); locally Lorentz invariant VSL
theories; theories exhibiting a color dependent speed of light;
varying $c$ induced by extra dimensions (e.g. in the brane-world
scenario); and field theories where VSL results from vacuum
polarization or CPT violation. We show how VSL scenarios may solve
the cosmological problems usually tackled by inflation, and also
how they may produce a scale-invariant spectrum of Gaussian
fluctuations, capable of explaining the WMAP data. We then review
the connection between VSL and theories of quantum gravity,
showing how ``doubly special'' relativity has emerged as a VSL
effective model of quantum space-time, with observational
implications for ultra high energy cosmic rays and gamma ray
bursts. Some recent work on the physics of ``black'' holes and
other compact objects in VSL theories is also described,
highlighting phenomena associated with spatial (as opposed to
temporal) variations in $c$. Finally we describe the observational
status of the theory. The evidence is currently slim -- redshift dependence
in the atomic fine structure, anomalies with
ultra high energy cosmic rays, and (to a much lesser
extent) the acceleration of the universe and the WMAP data. The
constraints (e.g. those arising from nucleosynthesis or geological
bounds) are tight, but not insurmountable. We conclude
with the observational predictions of the theory, and the prospects
for its refutation or vindication.

\vfill
\blankline

\blankline
j.magueijo@imperial.ac.uk
\eject
\tableofcontents
\eject
\vfill

\section{The black sheep of ``varying-constant'' theories}
\blankline

\blankline
\begin{center}
{\it One field of work in which there has been too much speculation is
cosmology. There are very few hard facts to go on, but theoretical
workers have been busy constructing various models for the universe,
based on any assumptions that they fancy. These models are probably
all wrong. It is usually assumed that the laws of nature have
always been the same as they are now. There is no justification for
this. The laws may be changing, and in particular quantities which
are considered to be constants of nature may be varying with cosmological
time. Such variations would completely upset the model makers.}
\blankline

Paul Dirac, ``On methods in theoretical physics'', June 1968, Trieste.
\end{center}
\blankline

\blankline

Since Dirac wrote these words in 1968, much has changed in our
understanding of the universe. It is fair to say that cosmologists
now have many ``hard facts to go on''. They have mapped the
cosmological expansion up to redshifts of order one
\cite{super,super1,super2,super3}. They have made high precision
observations of the cosmic microwave background
(CMB)~\cite{boom,wmap}, a major asset to observational cosmology
-- and hardly an established fact in 1968. Big Bang
nucleosynthesis has become a reasonably direct probe of the
conditions in the universe one second after the Big
Bang~\cite{bbn}. And there are many more: cosmologists can no
longer indulge in mere flights of fancy. Cosmology has finally
become an experimental science, or -- some might say -- a proper
science.

And yet, any statement about the universe's life before one second
of age is still necessarily a speculation. No observational
technique has so far penetrated this murky past, and Dirac's views
are still painfully applicable. In particular, it could well be
that the constants of nature are not constant at all, but were
varying significantly during this early phase. Assuming their
constancy at all times requires massive extrapolation, with no
observational basis. Could the universe come into being riding the
back of wildly varying constants?

As Dirac's quote shows, this question is far from new, and several
``constants'' of nature have been stripped off their status in
theories proposed in the past. Physicists have long entertained
the possibility of a varying gravitational constant
$G$~\cite{dirac,canuto,bdick},
%(e.g.
%putatively
%~\cite{dirac}, scale-covariant gravity~\cite{canuto},
%Brans-Dicke theories\cite{bdick}, or more generally scalar-tensor
%theories),
a varying electron charge $e$ \cite{bek2}, and more
generally varying coupling constants. Indeed with the advent of
string theory (and the prediction of the dilaton), to ``vary''
these ``constants'' seems to be fashionable.

In sharp contrast, the constancy of the speed of light has remain
sacred, and the term ``heresy'' is occasionally used in relation
to ``varying speed of light theories''~\cite{newsci}. The reason
is clear: the constancy of $c$, unlike the constancy of $G$ or
$e$, is the pillar of special relativity and thus of modern
physics. Varying $c$ theories are expected to cause much more
structural damage to physics' formalism than other varying
constant theories.

Ironically, the first ``varying-constant'' was the speed of light,
as suggested by Kelvin and Tait \cite{vsl19} in 1874. Some 30
years before Einstein's proposal of special relativity, a varying
c did not shock anyone, as indeed $c$ -- unlike $G$ -- played no
special role in the formalism of physics. This is to be contrasted
with the state of affairs  after 1905, when Eddington would say:
``A variation in c is self-contradictory''\cite{eddi}. This
astonishing statement does a disservice to the experimental
testability and scientific respectability of the theory of
relativity. In the words of Hertz, ``what is due to experiment may
always be rectified by experiment''.

In this review we describe how recent work has brought a varying
speed of light (VSL) into the arenas of cosmology,  quantum
gravity and experiment/observation. As a cosmological model, VSL
may be seen as a competitor to inflation, solving the cosmological
problems and providing a theory of structure formation. As a
theory of quantum gravity it may be seen as a phenomenological
project, more modest in scope than string theory or loop quantum
gravity, but already capable of making contact with experiment. On
the observational front it's still early days, but we could
already have seen evidence for VSL.

But despite these many-layered developments, some scientists still
question the logical consistency of varying the speed of light. It
seems befitting to start this review by addressing this matter.

\section{The meaning of a varying $c$ }\label{mean}
In discussing the physical meaning of a varying speed of light,
I'm afraid that Eddington's religious fervor is still with
us~\cite{ellis,duff}. ``To vary the speed of light is
self-contradictory'' has now been transmuted into ``asking whether
$c$ has varied over cosmic history is like asking whether the
number of liters to the gallon has varied''
\cite{duff}\footnote{I'd like to thank Mike Duff for persistently
disagreeing with me. However, splitting of hairs has been kept
out of the main text. One should not confuse the
constancy of the speed of light with its numerical value.
``Dimensionless'' and ``unit-invariant''
are the same thing when defined sensibly. }.
The implication is that the constancy of
the speed of light is a logical necessity, a definition that could
not have been otherwise. This has to be naive. For centuries the
constancy of the speed of light played no role in physics, and
presumably physics did not start being logically consistent in
1905. Furthermore, the postulate of the constancy of $c$ in
special relativity was prompted by experiments (including those
leading to Maxwell's theory) rather than issues of consistency.
History alone suggests that the constancy (or otherwise) of the
speed of light has to be more than a self-evident necessity.

\subsection{The argument against a varying $c$}

But let's examine the scientific merits of such a view. The
trouble arises because the attitude in \cite{dov,duff} (as opposed
to that in \cite{ellis}) is far from risible and is founded on a
perfectly correct remark~\footnote{At least in the case of
space-time variations of $c$; we'll examine the other cases
later.}. The speed of light is a quantity with units (units of
speed) and in a world without constants there is no a priori
guarantee that meter sticks are the same at all points and that
clocks spread throughout the universe are identical. Clearly if a
{\it dimensionless} constant is observed to vary -- such as the
fine structure constant $\alpha=e^2/\hbar c$ -- that fact is
unambiguous. But if $\alpha$ is seen to vary, the units employed
to quote physical measurements may also be expected to vary. A
meter stick may elongate or contract and a clock tick faster or
slower~\footnote{We are talking about clocks and rods at rest with
respect to the observer and far away from strong gravitational
fields.}. Hence under a changing $\alpha$ there is no guarantee
that units of length, time and mass are fixed, and discussing the
variability or constancy of a parameter with dimensions -- such as
the speed of light -- is necessarily circular and depends on the
definition of units one has employed. This remark was clearly made
by Bekenstein \cite{bek2}, who pointed out that the
``observation'' of a varying dimensional constant is at best a
tautology, since it relies on the definition of a system of units.
He stressed that the result of any experiment is necessarily
dimensionless, because it's the result of a ratio of two things
with the same dimensions: what is being measured and the ``unit''
employed. Hence, when assessing the constancy or variability of
constants, experiments are only sensitive to dimensionless
combinations of constants. From a strict  operational point of
view it only makes sense to talk about varying dimensionless
constants.

Seen from another angle~\cite{duff,ellis}, even in a world where
all seems to vary and nothing is constant, it is always possible
to {\it define} units such that $c$ remains a constant. Consider
for instance the current official definition of the meter: one
takes the period of light from a certain atomic transition as the
unit of time, then states that the meter is the distance travelled
by light in a certain number of such periods. With these
definitions it is clear that $c$ will always be a constant, a
statement akin to saying that the speed of light is one light-year
per year. One then does not need to perform any experiment to
prove the constancy of the speed of light: it is built into the
definition of the units and has become a tautology.

The argument is therefore double-tailed:
\begin{itemize}
\item  A
varying $c$ is tautological and is tied to the definition of a
system of units.
\item Units may always be defined so that $c$
becomes a constant.
\end{itemize}
It is not often pointed out that even though these arguments are
invariably invoked to attack a varying $c$, they apply equally
well to {\it any} other dimensional constant: $e$, $\hbar$, $G$,
etc. For exactly the same reasons one may argue that the
variability of these constants is tautological, or that units may
always be defined so that the variability envisaged by the theory
is undone. And yet varying $e$ and $G$ theories are widely
accepted. Are dilaton and Brans-Dicke theories
``self-contradictory'' as well? Taken without prejudice, the
arguments against a varying $c$ could destroy any varying constant
theory.

\subsection{The loophole}\label{loophole}

To fix ideas, we consider an observational example: Webb and
collaborators \cite{murphy,webb} have reported evidence for
redshift dependence in $\alpha$. Since $\alpha$ is dimensionless
it does not fall prey to the above arguments. But then the
question arises: given these observations, which of $e$, $\hbar$,
$c$, or a combination thereof is varying? Yet again, some authors
are keen to point out that interpreting the Webb et al results as
a varying $c$ is meaningless, due to the above arguments. They
fail to notice that when they state that these results are {\it
not} due to a varying $c$ (thus, being due to a varying $e$ or
$\hbar$) they are making an equally meaningless statement, for
exactly the same reasons ($e$ and $\hbar$ have dimensions).

If $\alpha$ is seen to vary one cannot say that all the
dimensional parameters that make it up are constant. Something --
$e$, $\hbar$, $c$, or a combination thereof -- has to be varying.
The choice amounts to fixing a system of units, but that choice
{\it has to be made}.

A possible way to evade this argument is to say that physical
theories should only refer to directly measurable dimensionless
parameters~\cite{duff}, a view I label fundamentalism. This
commendable view is, however, mere pub talk -- no one has ever set
up a theory in which only dimensionless parameters exist. At the
end of the day, even if all dimensionless parameters were running
wild, one would still want to set up quantum mechanics using
Planck's ``constant'', or electrodynamics using the speed of
light. Dimensional quantities would still play a role, something
made more obvious by noting that if ``constants'' do vary then
they're just quantities like any other (like fields, or the length
of my desk). And even under wildly varying constants, I would
still want to know the length of my desk in meters, no matter how
the meter is defined.

We need units and dimensional parameters to set up physics.
Dimensional parameters or quantities are a necessary evil in
physics. For the most part they are tautological and meaningless;
still within the whole construction one gleans operationally
meaningful statements, which are indeed dimensionless. But it's
easier to get there by means of constructions which are purely
human conventions. These conventions amount to a prescription for
defining units of mass, time, and length (and temperature if
required). In the context of varying dimensionless constants, that
choice translates into a statement on which dimensional constants
are varying.

\subsection{No subjectivism -- the example of Newtonian mechanics}

It would seem that we are  falling into subjectivism, but that is
not the case. The choice of units is never arbitrary or personal,
once one specifies a given dynamics (via a Lagrangian or
otherwise), which may then make predictions to be refuted or
verified by experiment. One system of units invariably renders the
presentation of the dynamics simpler than all others. Changing the
units would not change the physical content of the theory, but
would change its aspect. Typically one aspect is simple, and all
others are ridiculously complicated. This unambiguously fixes the
units to be used.

To give an example, there is a priori nothing wrong with using my
pulse as the unit of time, and rephrasing physics in
``egocentric'' units. The physical content of such a theory would
be the same --- but we know that the laws of physics would {\it
look} pretty weird, while being operationally the same. There
would even be the illusion of seemingly new phenomena: for
instance, every time I ran to catch a bus the speed of light would
decrease. But nothing would be physically different, and according
to the fundamentalist view, this description would be perfectly
acceptable, or at least as meaningless as the conventional
description.

Less preposterously, there was once a time when one might be
tempted to define time by means of a pendulum, and rephrase
Newtonian gravity by insisting that a pendulum clock is the
``right'' way to keep time. According to such a ``pendular
physics'', objects would be more rigid and the speed of light
higher for observers on the Moon (since a pendulum clock would
tick slower than the conventional Newtonian clocks). Newton's laws
would {\it look} much more complicated (except for the $1/r^2$
law), but the physical content of the theory would be the same.
And yet Newton did not do this: he picked a more sensible system
of units, one which rendered the law of inertia, the uniformity of
time, and the conservation of energy valid.

As these examples show, the ability to formulate a theory (in this
case Newtonian mechanics) often depends on choosing the right
units. This is far from new, and has been discussed at length in
the past. To cite Poincar\'e \cite{poincare}: ``If now it be
supposed that another way of measuring time is adopted, the
experiments of which Newton's law is founded would nonetheless
have the same meaning. Only the enunciation of the law would be
different (...). So that the definition implicitly adopted (...)
may be summed up thus: time should be so defined that the
equations of mechanics may be as simple as possible. In other
words, there is not one way of measuring time more true than
another; that which is adopted is only more {\it convenient}.''

The implications of this statement are far reaching. Poincar\'e
clearly implies that matters as fundamental as the uniformity of
time, and by consequence the law of inertia and the theorem of
energy conservation, are not provable by experiment. Experiment is
dimensionless, but these statements ``have units'', e.g. depend on
the definition of the unit of time, which is nothing but a
convention. And yet that definition is not a subjective choice.
One particular unit of time -- that which renders the laws of
classical mechanics simple -- {\it objectively} stands out.

%Indeed a simple way to construct such a clock is to take a free
%object, a meter stick, and use the time it takes to cover on meter
%as a clock (a definition not dissimilar to the definition of the
%meter in the post-relativity days). But it tells you more: it
%tells you to do so because you will then be able to express with
%great simplicity the whole framework of Newtonian physics.

\subsection{A general definition of VSL}\label{defvsl}

The above discussion has many parallels with VSL, as one final
example shows.
% once one notes that the constant speed which
%appears in law inertia
%--- objects not acted upon (read, far from everything) move at a
%constant speed
%--- is a tautology, according to the fundamentalist view. The
%constant speed in the law of inertia is as tautological as the
%constant speed of light. The law of inertia merely tells you that
%you can use a clock such that the law of inertia is valid.
%
%To narrow down on VSL, consider one final example: i
In classical electromagnetism the speed of light is only constant
in vacuum, and it ``varies'' in dielectric media. This statement
falls prey to all the criticism usually directed at VSL, namely
that we could choose units such that the speed of light {\it in
dielectric media} is a constant. As in the example given by
Poincar\'e, no doubt you could do that; however such a convention
would render the enunciation of Maxwell's laws in dielectric media
very complicated. Instead of simply replacing $c$ by
$c_0/\sqrt\epsilon$, in ``constant $c$'' units one would need to
add new terms in gradients and time derivatives of $\epsilon$ to
Maxwell's equations. Simplicity tells you that in this context you
should not {\it choose} units in which the speed of light is a
constant.

We are now ready to define varying speed of light. VSL theories
are theories in which you find yourself in a situation like the
one in the last example, regarding the speed of light {\it in
vacuum}. They are theories in which the dynamics is rendered more
simple if units are chosen in which $c$ is not constant. Typically
this can be achieved if Lorentz invariance is broken, or if the
usual tools employed in differential geometry become frame
dependent. But the point is that we cannot discuss ``VSL vs
constant $c$'' until a specific dynamics is proposed. One may then
discover that varying $c$ units are preferable: in
Section~\ref{cosmoeqs} we will give an explicit example (compare
Eqns~(\ref{fried1}) and (\ref{fried2}) with Eqns~(\ref{fried1p})
and (\ref{fried2p})).

To return to the issue of the meaning of the observed varying
$\alpha$, we note that while observers concern themselves with
dimensionless quantities, theorists need dimensional quantities to
set up their theories.  In order to set up a theory it may be more
convenient to choose one system of units rather than any other. In
dilaton theories, or variants thereof
\cite{bsbm,bek2,olive,uzan,revmartins}, the observed variations in
$\alpha$ are attributed to $e$; VSL
theories~\cite{moffat93,am,ba,bm,barmag98,bdicke,covvsl,peres}
blame $c$ for this variation (and in some cases $\hbar$ too, see
\cite{covvsl}). These choices are purely a matter of convenience,
and one may change the units so as to convert a VSL theory into a
constant $c$, varying $e$ theory; however such an operation is
typically very contrived, with the resulting theory looking
extremely complicated. Hence the dynamics associated with each
varying $\alpha$ theory ``chooses'' the units to be used, on the
grounds of convenience, and this choice fixes which combination of
$e$, $c$ and $\hbar$ is {\it assumed} to vary.

The good news for experimentalists is that once this theoretical
choice is made, the different theories typically lead to very
different predictions. Dilaton theories, for instance, violate the
weak equivalence principle, whereas many VSL theories do
not~\cite{mofwep,mbswep}. VSL theories often entail breaking
Lorentz invariance, whereas dilaton theories do not. These
differences have clear observational implications, for instance
the STEP satellite could soon rule out the dilaton theories
capable of explaining the Webb et al results~\cite{stepsat}.
Violations of Lorentz invariance, as we shall see, should also
soon be observed - or not. We shall return to these matters in
Section~\ref{obstat}.

\subsection{Dimensionless varying $c$}
We conclude by noting that the above considerations apply to
theories displaying space-time variations in the speed of light,
and that there are theories for which a varying $c$ is a
dimensionless statement~\footnote{This does not include
\cite{davis}, where a dimensionless statement on varying $c$ is
achieved only because a constant $G$ was assumed.}. For instance
$c$ may be color dependent (e.g. \cite{camreview}). One may then
take two light rays, measure their frequencies and speeds (in
whatever units) and compute the ratios between the two frequencies
and between the two speeds. Both ratios are dimensionless. If the
latter is different from one when the former is also different
from one, we have an example of varying speed of light which does
not depend on the units employed. Another example is bimetric
VSL~\cite{mofclay,drum}, for which the speed of the photon and
graviton may differ. One may form the ratio between the speed of
the photon and that of the graviton, to form a dimensionless
quantity associated with a varying speed of light.~\footnote{
Truly paranoid physicists may argue that there is no
guarantee that the units employed remain the same at different
energies, or when measuring gravitons instead of photons. In this
very extreme sense there can be no dimensionless varying constants.
}

\section{An abridged catalogue of recent VSL theories}
Even after the proposal of special relativity in 1905 many varying
speed of light theories were considered, most notably by Einstein
himself~\cite{einsvsl}. VSL was then rediscovered and forgotten on
several occasions. For instance, in the 1930s VSL was used as an
alternative explanation for the cosmological redshift
\cite{stew,buc,wald} (these theories conflict with fine structure
observations). None of these efforts relates to recent VSL
theories, which are firmly entrenched in the successes (and
remaining failures) of the hot big bang theory of the universe. In
this sense the first ``modern'' VSL theory was J. W. Moffat's ground
breaking paper \cite{moffat93}, where spontaneous symmetry
breaking of Lorentz symmetry leads to VSL and an elegant solution
to the horizon problem.

Since then there has been a growing literature on the subject,
with several groups working on different aspects of VSL (for an
early review see~\cite{kish}).
%--- work that renders my early turn of phrase ``black sheep''
%a bit of a joke.
In this section I categorize the main implementations
currently being considered, without trying to be exhaustive.
All VSL theories conflict in one
way or another with special relativity and here I shall use the
type of insult directed at special relativity as my classification
criterion for VSL theories.

Recall that special relativity is based upon two {\it independent}
postulates - the relative nature of motion and the constancy of
the speed of light. VSL theories do not need to violate the first
of these postulates, but in practice one finds it hard to dispense
with the second without destroying the first. This leads to our
first criterion for differentiating the various proposals: do they
honor the relative nature of motion?

Regarding the second postulate of special relativity, VSL theories
behave in a variety of ways, all arising from a careful reading of
the small print associated the constancy of $c$. Loosely this
postulate means that $c$ is a constant, but more precisely it
states that the speed of all {\it massless} particles is the same,
regardless of their color (frequency), direction of motion, place
and time, and regardless of the state of motion of observer or
emitter. There is a large number of combinations in which these
different aspects can be violated, explaining the large number of
VSL theories put forward\footnote{Each of these theories gives
a detailed answer to the questions raised in \cite{alluz}. The
authors of \cite{alluz} are painfully ignorant of the ins and outs
of the various VSL proposals.}.

Bearing this in mind we can now distinguish the following VSL
mechanisms.

\subsection{Hard breaking of Lorentz symmetry}\label{hardvsl}
The most extreme model is that proposed in \cite{am}, and studied
further in \cite{ba}. In this model both postulates of special
relativity are violated: there is a preferred frame in physics
(usually identified with the cosmological frame); the speed of
light varies in time, although usually only in the very early
Universe; and the time-translation invariance of physics is
broken~\cite{wheeler}. It describes a world where not only the
matter content of the universe, but also the laws of physics
evolve in time.

The basic dynamical postulate is that
Einstein's field equations are valid, with minimal coupling (i.e.
with $c$ replaced by a field in the relevant equations) in this
particular frame:
\begin{equation}
G_{\mu \nu }-g_{\mu \nu }\Lambda ={\frac{8\pi G}{c^4} }T_{\mu \nu
}.
\end{equation}
This is inspired by the statement of Maxwell's equations in
dielectric media. The first strong assumption is that the field
$c$ does not contribute to the stress-energy tensor (this may also
be implemented in more conservative theories, e.g.~\cite{covvsl}).
More importantly, this postulate can only be true in one frame.
There is still  a metric, a connection, and curvature and Einstein
``tensors'', to be evaluated in a given frame {\it at constant
$c$} (no extra terms in gradients of $c$, e.g. in the expression
for the connection in terms of the metric). But they are tied to a
preferred frame, and non-covariant extra terms in gradients of $c$
will appear in other frames. As pointed out by \cite{bass} this is
no longer a geometric theory. Minimal coupling at the level of
Einstein's equations is at the heart of the model's ability to
solve the cosmological problems (as we shall see in
Section~\ref{cosmo}).

It is debatable whether such a theory may derive from an action
principle formulation. We could perform all variations at constant
$c$ and require that  ${\cal L}_c $ must not contain the metric
explicitly.  Unsurprisingly a hamiltonian formulation of this
theory is preferable \cite{barmag98}. The preferred frame is given
by a 4-vector $u^\mu $, and the metric can be written as
\begin{equation}
g_{\mu \nu }= h_{\mu \nu} - u_{\mu} u_{\nu}
\end{equation}
with $h_{\mu \nu }u^\mu =0$ and $u_\mu ^\mu =-1$. In a
cosmological setting this frame is defined by the proper time and
the conformal space coordinates which ensure that $K=0,\pm 1$. The
Einstein equations derived from this action are more simply
written using the Hamiltonian formalism, with a 3+1 split induced
by vector $u^\mu $ (see for instance \cite{mtw}). With a unit
lapse and zero shift (temporal gauge), the Hamiltonian density in
Einstein's theory takes the form
\begin{equation}
{\cal H}=h^{1/2}{\left[ -^{(3)}R+h^{-1}{\left( \Pi ^{ij }\Pi _{ij
}-{\frac 12}\Pi ^2\right) }\right] }
\end{equation}
(with $i,j=1,2,3$).
The second fundamental form is given by
\begin{equation}
\kappa _{ij }={1\over c}{\frac{{\dot h_{ij }}}2}
\end{equation}
and the momenta conjugate to the $h_{ij }$ are given by
\begin{equation}
\Pi _{ij }={\frac{\partial {\cal L}}{\partial {\dot h}^{ij }}}%
=h^{1/2}(\kappa _{ij }-\kappa h_{ij}).
\end{equation}
The Hamiltonian constraint is ${\cal H}=0$ and the momentum constraint is
\begin{equation}
\nabla _i (h^{-1/2}\Pi ^{ij })=0
\end{equation}
The dynamical equations are
\begin{equation}
{1\over c}{\dot h}_{ij }
%={\frac{\delta {\cal H}}{\delta \Pi ^{ij }}}%
=2h^{-1/2}{\left( \Pi _{ij }-{\frac 12}h_{ij }\Pi \right) }
\end{equation}
and
\begin{eqnarray}
{1\over c^2}{{\sqrt h}\over 2}[\dot h_{ij}-\dot h h_{ij}]\, \dot{ }&=
%&\nonumber \\
%-{\frac{\delta {\cal H}}{\delta h^{\mu \nu }}}\nonumber \\
%&=
&-h^{1/2}  {\left( ^{(3)}R_{ij }-{\frac 12}h_{ij }\,
^{(3)}R\right) } \nonumber \\
&&\ +{\frac 12}h^{-1/2}h_{ij }{\left( \Pi _{kl }\Pi ^{kl
}-{\frac 12}\Pi ^2\right) }  \nonumber \\
&&\ -2h^{-1/2}{\left( \Pi _i ^k \Pi _{kj }-{\frac 12}\Pi \Pi
_{ij }\right) }
\end{eqnarray}
These are Einstein's
equations in vacuum, but an adaptation to include matter is easy
to write down.
 These equations differ considerably when written in
units in which $c$ does not vary, losing their minimal coupling
aspect~\cite{barmag98}. This is reminiscent of Maxwell's equations
in dielectric media if rewritten in units in which the speed of
light in these media is a constant (we will give an example in
Section~\ref{cosmoeqs}). They also look considerably different in
a different frame, signaling the breakdown of covariance. Indeed
it is easy to show that these theories do not satisfy Bianchi
identities \cite{bass}. Although in the initial model the speed of
light, like the preferred frame $u^\mu$, was preset (and thus to
be seen as a law of physics), this need not be the case. In
Section~\ref{cosmo} we shall show how a dynamical equation for $c$
may also be included in this formalism.

Other theories with a preferred frame $u^\mu$ have been considered
in the past, e.g. the ``aether theory'' of \cite{jacobson}, or the
CPT-odd theory of Coleman and Glashow~\cite{colglash}
(check out the 4-vector $s^\mu$ in their Eqn. (6)).

\subsection {Bimetric VSL theories}\label{bivsl}
This approach was initially proposed by J. W. Moffat and M. A. Clayton
\cite{mofclay}, and by I. Drummond \cite{drum}. In contrast with
the above formulation, one does not sacrifice the first principle
of special relativity and special care is taken with the damage
caused to the second. In these theories the speeds of the various
massless species may be different, but special relativity is still
realized within each sector. Typically the speed of the graviton
is taken to be different from that of massless matter particles.
This is implemented by introducing two metrics (or tetrads in the
formalism of \cite{drum,drum1}), one for gravity and one for
matter. The model was further studied by \cite{clay1}
(scalar-tensor model), \cite{clay} (vector model), and
\cite{covvsl,bass,bass1}.

We now sketch an implementation of the scalar-tensor model. It
uses a scalar field $\phi$ that is minimally coupled to a
gravitational field described by the metric $g_{\mu\nu}$. However
the matter couples to a different metric, given by:
\begin{equation}
\hat{g}_{\mu\nu}=g_{\mu\nu}+B\partial_\mu\phi\partial_\nu\phi.
\end{equation}
Thus there is a space-time, or graviton metric $g_{\mu\nu}$, and a
matter metric $\hat{g}_{\mu\nu}$. The total action is:
\begin{equation}
S=S_g+S_{\phi}+\hat{S}_{\rm M},
\end{equation}
where the gravitational action is as usual
\begin{equation}
S_g=-\frac{c^4}{16\pi G}\int dx^4{\sqrt {-g}} (R(g)+2\Lambda),
\end{equation}
(notice that the cosmological constant $\Lambda$ could also,
non-equivalently, appear as part of the matter action). The scalar
field action is:
\begin{equation}
S_\phi=\frac{c^4}{16\pi G}\int dx^4{\sqrt {-g}}\,
\Bigl[\frac{1}{2}g^{\mu\nu}\partial_\mu\phi\partial_\nu\phi-V(\phi)\Bigr],
\end{equation}
leading to the the stress-energy tensor:
\begin{equation}
T^{\mu\nu}_\phi=  \frac{c^4}{16\pi G} \Bigl[
g^{\mu\alpha}g^{\nu\beta}\partial_\alpha\phi\partial_\beta\phi
-\frac{1}{2}g^{\mu\nu}g^{\alpha\beta}\partial_\alpha\phi\partial_\beta\phi
+g^{\mu\nu}V(\phi) \Bigr].
\end{equation}
The matter action is then written as usual, but using the metric
${\hat g}_{\mu\nu}$. Variation with respect to $g_{\mu\nu}$ leads
to the gravitational field equations:
\begin{equation}
G^{\mu\nu}=\Lambda g^{\mu\nu}
 +\frac{8\pi G}{c^4}   T^{\mu\nu}_\phi
 +\frac{8\pi G}{c^4}    \frac{\sqrt{-\hat{g}}}{\sqrt {-g}}\hat{T}^{\mu\nu}.
\end{equation}
In this theory the speed of light is not preset, but becomes a
dynamical variable predicted by a special wave equation
\begin{equation}\label{dynabivsl}
\bar{g}^{\mu\nu}\hat{\nabla}_\mu\hat{\nabla}_\nu\phi+KV^\prime
[\phi]=0\,
\end{equation}
where the biscalar metric $\bar g$ is defined in \cite{mofclay}.

This model not only predicts a varying speed of light (if the
speed of the graviton is assumed to be constant), but also allows
solutions with a de Sitter phase that provides sufficient
inflation to solve the horizon and flatness problems. This is
achieved without the addition of a potential for the scalar field.
The model has also been used as an alternative explanation for the
dark matter \cite{drum1} and dark
energy\cite{bass,bass1}. In Section~\ref{struc} we shall describe
the implications of this model for structure formation.

\subsection{Color-dependent speed of light}\label{colvsl}
This approach may or may not preserve the first postulate of
special relativity, the relative nature of motion; however,
it generally violates its second postulate in the sense that the
speed of light is allowed to vary with color (typically only
close to the Planck frequency).
This is achieved by deforming the photon dispersion relations
$E^2-p^2=m^2=0$. For instance, it was proposed that:
\begin{equation}\label{dispcam}
    E^2 = p^2 + m^2 + \lambda E^3 + ...
\end{equation}
where $\lambda$ is of the order of the Planck length. If this
dispersion relation is true in one frame, and if the linear
Lorentz transformations are still valid, then they are not true in
any other frame, and so this theory -- like that of \cite{am}
described in Section~\ref{hardvsl} -- contradicts the principle of
relativity. This is the case in the  pioneering work
in~\cite{amel,amel1,liouv} where the effect was found for
space-time foam. However this need not be the
case~\cite{amelstat,gli,leejoao}. One may preserve full Lorentz invariance
(with a non-linear realization) and still accommodate deformed
dispersion relations.

Under deformed dispersion relations, the group velocity of light,
$c=dE/dp$, acquires an energy dependence. Such a dependence could
be observed in planned gamma ray observations \cite{amel}. An
energy dependent speed of light may also imply that the speed of
light was faster in the very early universe, when the average
energy was comparable to Planck energies~\cite{ncvsl}. This may
be used to implement VSL cosmology and even inflation~\cite{ncinfl}.
Such theories make viable predictions for structure
formation~\cite{pogo}, as explained further in
Section~\ref{struc}. Finally a modified dispersion relation may
lead to an explanation of the dark energy, in terms of energy
trapped in very high momentum and low-energy quanta, as pointed
out by~\cite{laura}.

But these theories are popular mainly as phenomenological
descriptions of quantum gravity (e.g.~\cite{leejoao,leejoao1}) and
as an explanation for the threshold anomalies
(e.g.~\cite{cosmicray,leejoao1}). We will describe
these aspects in more detail in Section~\ref{qg}.

\subsection{``Lorentz invariant'' VSL theories}\label{livsl}
It is also possible to preserve the essence of Lorentz invariance
in its totality and still have a space-time (as opposed to energy
dependent) varying $c$. One possibility is that Lorentz invariance
is spontaneously broken, as proposed by J. W. Moffat in his seminal
paper \cite{moffat93,moff2} (see also \cite{jacobson}). Here the
full theory is endowed with exact local Lorentz symmetry; however
the vacuum fails to exhibit this symmetry. For example an $O(3,1)$
scalar field $\phi^a$ (with $a=0,1,2,3$) could acquire a time-like
vacuum expectation value (vev), providing a preferred frame and
spontaneously breaking local Lorentz invariance to $O(3)$
(rotational invariance). Such a vev would act as the preferred
vector $u^a=\phi^a_0$ used in Section~\ref{hardvsl}; however the
full theory would still be locally Lorentz invariant. Typically in
this scenario  the speed of light undergoes a first or second
order phase transition to a value more than 30 orders of magnitude
smaller, corresponding to the presently measured speed of light.
Interestingly, before the phase transition the entropy of the
universe is reduced by many orders of magnitude, but afterwards
the radiation density and entropy of the universe vastly increase.
Thus the entropy increase follows the arrow of time determined by
the spontaneously broken direction of the timelike vev $\phi^a_0$.
This solves the enigma of the arrow of time and the second law of
thermodynamics (we will discuss the implications of this work for
quantum cosmology later).

Another example is the covariant and locally Lorentz invariant
theory proposed in \cite{covvsl}. In that work definitions were
proposed for covariance and local Lorentz invariance that remain
applicable when the speed of light $c$ is allowed to vary. They
have the merit of retaining only those aspects of the usual
definitions which are invariant under unit transformations, and
which can therefore legitimately represent the outcome of an
experiment (see discussion in Section~\ref{mean} above). In the
simplest case a scalar field is then defined $\psi=\log( c/ c_0)$,
and minimal coupling to matter requires that \be\label{alphan}
\alpha_i\propto g_i\propto \hbar c\propto c^{q} \ee  with $q$ a
parameter of the theory. The action may be taken to be
\begin{equation} S= \int d^4x \sqrt{-g}(
e^{a\psi}( R-2\Lambda +{\cal L}_{\psi}) +{ 16\pi G\over
c_0^4}e^{b\psi}{\cal L}_m )
\end{equation}
and the simplest dynamics for $\psi$ derives from:
\begin{equation} {\cal L}_{\psi}=-\kappa(\psi)
\nabla_\mu\psi\nabla^\mu\psi
\end{equation}
where $\kappa(\psi)$ is a dimensionless coupling function. For
$a=4$, $b=0$, this theory is nothing but a unit transformation
applied to Brans-Dicke theory. More generally, it's only when
$b+q=0$ that these theories are scalar-tensor theories in
disguise. In all other cases it has been shown that a unit
transformation may always be found such that $c$ is a constant but
then the dynamics of the theory becomes much more complicated (see
discussion in Section~\ref{mean}).

In these theories the cosmological constant $\Lambda$ may depend
on $c$, and so act as a potential driving $\psi$. Since the vacuum
energy usually scales like $c^4$ we may take $\Lambda\propto
(c/c_0)^n=e^{n\psi}$ with $n$ an integer. In this case, if we set
$a=b=0$ the dynamical equation for $\psi$ is :
\be\label{dynacovvsl} \Box \psi ={32\pi G\over c^4\kappa}{\cal
L}_m +{1\over \kappa} n\Lambda \ee Thus it is possible, in such
theories, that the presence of Lambda drives changes in the speed
of light, a matter examined (in another context) in \cite{vslsn}.

Particle production and second quantization for this model has
been discussed in \cite{covvsl}. Black hole solutions were also
extensively studied \cite{vslbh}, and some results will be
summarized in Section~\ref{bh}. Predictions for the classical
tests of relativity (gravitational light deflection, gravitational
redshift, radar echo delay, and the precession of the perihelion
of Mercury) were also shown to differ distinctly from their
Einstein counterparts, while still evading experimental
constraints~\cite{vslbh}. Other interesting results were the
discovery of Fock-Lorentz space-time\cite{man,step} as the
``free'' solution, and fast-tracks (tubes where the speed of light
is much higher) as solutions driven by cosmic
strings~\cite{covvsl}.

Beautiful as these two theories may be, their application to
cosmology is somewhat cumbersome.

\subsection{String/M-theory efforts}
This line of VSL work was initiated by Kiritis \cite{kir} and
Alexander \cite{steph}, and makes use of the brane-world scenario,
in which our Universe is confined to a 3-brane living in a space
with a larger number of dimensions (the bulk). They found that if
the brane lives in the vicinity of a black hole it is possible to
have exact Lorentz invariance (and hence a constant speed of light
in the bulk) while realizing VSL {\it on the brane}. In this
approach VSL results from a projection effect, and the Lorentz
invariance of the full theory remains unaffected.

More specifically it is found that the first order kinetic terms of
the gauge fields living in the brane are of the form:
\begin{equation}
S_{GF} = 2 T_{3} \int d^{4} x {\left( \frac{1}{\sqrt {f(r)}}\vec{ E}^{2}
+ \sqrt {f(r)} \vec{B}^{2} \right)} \, .
\end{equation}
where $\vec{ E}$ and $\vec{ B}$ are the electric and the magnetic
fields. Thus the speed of light depends on the distance $r$
between the brane and the black hole: \be
c(r)=c_0\sqrt{f(r)}=c_0\sqrt{1-{\left(\frac{r_0}{r}\right)}^{4}} \, .
\ee It decreases as the probe-brane universe approaches the black
hole and vanishes at the horizon.

Several posterior realizations of VSL from extra dimensions  make
use of the Randall-Sundrum models~\cite{rs1,rs2}, in which the
extra dimensions are subject to warped compactification. It was
shown in \cite{kal,frees,ishi} that light signals in such
space-times may travel faster through the extra dimensions. This
clearly does not conflict with relativity, where the ``constant
$c$'' always refers to the local -- and not the global -- speed of
light (for instance the global speed of light in a radiation
dominated Friedmann model is $2c$). However, when projected upon
the brane world it could appear that the local speed of light is
not constant. Further aspects of violations of Lorentz violations
for the projected physics on the brane were studied in
\cite{csaki,you1,quev}.

In particular Youm did extensive work showing how some of the
previously proposed VSL models may be realised in the brane-world
scenario. In \cite{you1,you2} he showed how to implement the model
of \cite{am} (see Section~\ref{hardvsl}) using a Randall-Sundrum
type model. However he found these models more restrictive,
regarding the ``desirable VSL'', than those emerging from standard
general relativity. Nonetheless, in the opposite direction, he
found that VSL could provide a possible mechanism for controlling
quantum corrections to the fine-tuned brane tensions after the
SUSY breaking. He also showed~\cite{doubime} how the bimetric VSL
model (see Section~\ref{bivsl}) could be realized in similar
circumstances, with the ``biscalar'' field assumed to be confined
on the brane. A model with a varying electric charge and more
generally the implications for varying alpha were also considered
\cite{youa1,alphadou}.

Still related to the Randall-Sundrum model there is mirage
cosmology (e.g. \cite{mirage}). In these models the brane motion
through the bulk may induce Friedmann cosmology on the brane even
if no matter is confined to it (hence the term ``mirage''). In
\cite{steer} a brane is considered embedded in two specific 10
dimensional bulk space-times: Sch-AdS$_5 \times$S$_5$ and a
rotating black hole. The projected Friedmann equations are then
found, the ``dark fluid'' terms are identified, and a varying
speed of light effect studied. It is found that the effective
speed of light in these models always increases -- hardly what is
desirable for cosmological applications.

Finally it should be stressed that some of the work
\cite{ncvsl,ncinfl,niem,heis} on deformed dispersion relations,
associated with a color dependent speed of light (see
Section~\ref{colvsl}) was inspired by non-commutative
geometry~\cite{snyder,jackiw,castorina}, which in turn appears to
be related to string/M theory~\cite{seib}. A further connection
relates to Liouville strings \cite{liouv,mavro,mavro1}. An unusual
example of VSL derived from an extra dimension can be found in
\cite{vone}.

In a different direction one may speculate how string theory would
react to a varying $c$ imposed at a fundamental level (i.e. {\it
not} as a projection effect). Some preliminary work shows that at
least for the bosonic string VSL would be disastrous, leading to
an energy dependent critical dimension~\cite{ljstring}. However
this result is far from general, and in the absence of
``M-theory'' the question remains unanswered.

J. W. Moffat has also pointed out~\cite{smatrix} that VSL could solve
the usual problem posed for string theory and quantum field
theories by the existence of future horizons. In particular, the
accelerating universe would appear to rule out the formulation of
physical S-matrix observables. Moffat remarked that postulating
that the speed of light varies in an expanding universe in the
future as well as in the past can eliminate future horizons,
allowing for a consistent definition of S-matrix observables, and
thus of string theory.

\subsection{Field theory VSL predictions}
It has been known for a while that quantum field theory in curved
space-time predicts superluminal photon propagation. This was
first pointed out in \cite{drhath}, where one-loop vacuum
polarization corrections to the photon propagation were computed
in a variety of backgrounds. One can in general find directions of
motion, or polarizations, for which the photon moves ``faster than
light''. Further examples can be found in \cite{shore,tey}.
Typically one distinguishes between the $c$ appearing in the
Lorentz transformations and the actual velocity of light, modified
due to non-minimal coupling to gravity. An early solution of the
horizon problem by means of this effect may be found
in~\cite{novello}. The implications for optics and causality of
this ``faster than light'' motion are discussed further in
\cite{shoreopt}.

The Casimir effect is another example where  VSL has been
discovered in field theories. As shown in \cite{scharn}, vacuum
quantum effects induce an anomalous speed of propagation for
photons moving perpendicular to a pair of conducting plates.

Regarding explicit breaking of Lorentz invariance (and not just a
varying $c$) it is also possible that Lorentz invariance is a low
energy limit. Indeed the work of Nielsen  and collaborators
suggests that Lorentz invariance could be a stable infrared fixed
point of the renormalization group flow of a quantum field theory
\cite{nielsen}.

On a different front it is known that violating Lorentz invariance
and breaking CPT are very closely related - the so-called CPT
theorem, where CPT invariance is deduced from Lorentz invariance
and locality alone. See for instance the recent result
\cite{greenberg} showing that CPT violation requires violations of
Lorentz invariance. Thus high energy physics tests of CPT can also
act as tests of Lorentz invariance~\cite{coleman,kosta,bert}, and
VSL may be studied in the framework of Lorentz violating extensions
of the standard model.

Neutrino oscillations are another upcoming area in this respect.
Flavor eigenstates do not satisfy standard dispersion relations
and may thus be related to the work described in
Section~\ref{colvsl}. Implications for the endpoint of beta decay
are currently being studied~\cite{neut}.

\subsection{Hybrids}
It is important to stress that the above examples are in general
not mutually exclusive.  For instance in \cite{burg} one may find
a brane-world scenario leading to graviton deformed dispersion
relations and bimetric VSL. Also, one could easily overlay
space-time variations and color dependence in the speed of light,
as first pointed out in \cite{ncvsl}. The low energy $c_0$ which
appears in all work on color-dependent speed of light
(Section~\ref{colvsl}), could itself be a space-time variable
identified with the varying speed of light considered in the
models in Sections~\ref{hardvsl},~\ref{bivsl}, and \ref{livsl}.

One may argue that such combinations are baroque, but this is an
aesthetical prejudice which should be set aside when discussing
the observational status of VSL (Section~\ref{obstat}).

\section{Varying-$c$ solutions to the Big Bang problems}\label{cosmo}
Like inflation~\cite{infl}, modern VSL theories were  motivated by
the ``cosmological problems'' -- the flatness, entropy,
homogeneity, isotropy  and cosmological constant problems of Big
Bang cosmology (see~\cite{am,basker} for a review, and
\cite{numbers} for a dimensionless description). The definition
of a cosmological arrow of time was also a strong consideration.

\begin{figure}
\centerline{\psfig{file=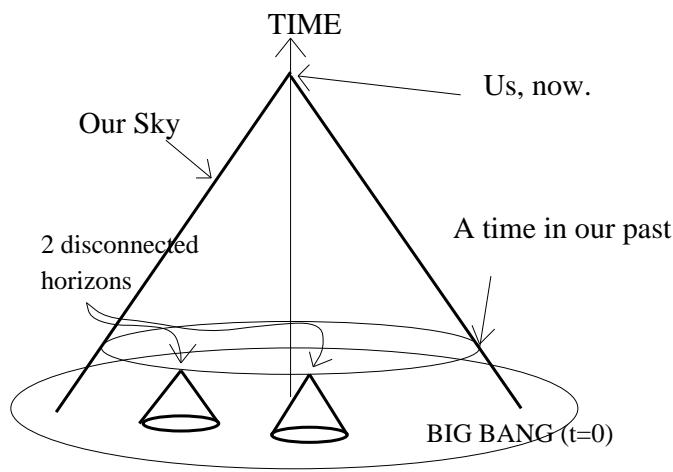,width=6 cm,angle=-90}}
\caption{A conformal diagram (in which light travels at $45^\circ$).
This diagram reveals that the sky is a cone in 4-dimensional
space-time. When we look far away we look into the past; there is
an horizon because we can only look as far away as the Universe is
old. The fact that the horizon is very small in the very early
Universe, means that we can now see regions in our sky outside
each others' horizon. This is the horizon problem of standard Big
Bang cosmology.} \label{fig1}
\end{figure}

\begin{figure}
\centerline{\psfig{file=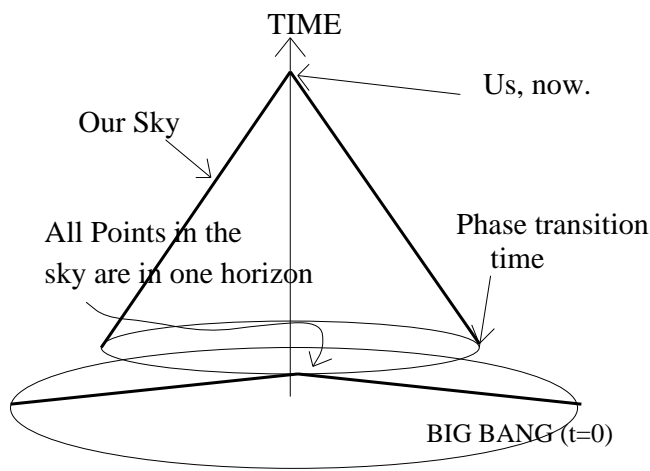,width=6 cm,angle=-90}}
\caption{Diagram showing the horizon structure in a model in which
at time $t_c$ the speed of light changed from $c^-$ to $c^+\ll
c^-$. Light travels at $45^\circ$ after $t_c$ but it travels at a
much smaller angle to the spatial axis before $t_c$. Hence it is
possible for the horizon at $t_c$ to be much larger than the
portion of the Universe at $t_c$ intersecting our past light cone.
All regions in our past have then always been in causal contact.
This is the VSL solution of the horizon problem.} \label{fig2}
\end{figure}

But at its most simplistic, VSL was inspired by the horizon problem.
As we go back into our past the present comoving horizon breaks down
into more and more comoving causally connected regions. These
disconnected early days of the Universe prevent a physical
explanation for the large scale features we observe -- the ``horizon
problem''. It does
not take much to see that a larger speed of light in the early
universe could open up the horizons~\cite{moffat93,am} (see
Figs.~\ref{fig1} and \ref{fig2}). More mathematically,
the comoving horizon is given by
$r_h=c/\dot a$, so that a solution to the horizon problem requires
that in our past $r_h$ must have decreased in order to causally
connect the large region we can see nowadays. Thus
\begin{equation}
{\ddot a\over \dot a}-{\dot c\over c}>0
\end{equation}
that is, either we have accelerated expansion (inflation), or a
decreasing speed of light, or a combination of both. This argument
is far from general: a contraction period ($\dot a<0$, as in the
bouncing universe), or a static start for the universe ($\dot
a=0$) are examples of exceptions to this rule.

However the horizon problem is just a warm up for the other
problems. We now illustrate how VSL can solve these problems using
as an example the model of \cite{am} (the solution in other models
is often qualitatively very similar).

\subsection{Cosmology and preferred frames}\label{preframe}
The VSL cosmological model first proposed in \cite{am} unashamedly
makes use of a preferred frame, thereby violating the principle of
relativity. In this respect a few remarks are in order. Physicists
don't like preferred frames, but they often ignore the very
obvious fact that we {\it have} a great candidate for a preferred
frame: the cosmological frame. This frame is a witness to all our
experiments -- we have never performed an experiment without the
rest of the Universe being out there -- a fact first pointed out
by Mach in relation to what he called the ``fixed stars''.

A complication, however, arises because
we are in motion with respect to this frame, as revealed
by the CMB dipole. This dipole has never been seen to permeate
the laws of physics. In addition, every six months our motion
around the Sun adds or subtracts a velocity with respect to the
cosmological frame and we fail to see corresponding fluctuations in laboratory
physics. The witness is therefore not very talkative, and if the
laws of physics are indeed tied to the cosmological frame, its
direct influence upon them has to be subtle.

Modern physicists invariably {\it choose} to formulate their laws
without reference to this preferred frame. This is more due to
mathematical or aesthetical reasons than anything else: covariance
and background independence have been regarded as highly cherished
mathematical assets since the proposal of general relativity. The
VSL theory proposed in \cite{am} makes a radically different
choice in this respect: it ties the formulation of the physical
laws to the cosmological frame. The basic postulate is that
Einstein's field equations are valid, with minimal coupling (i.e.
with $c$ replaced by a field in the relevant equations) in this
particular frame. This can only be true in one frame; thus,
although the dynamics of Einstein's gravity is preserved to a
large  extent, the theory is no longer geometrical. However, if
$c$ does not vary by much, the effects of the preferred frame are
negligible, of the order $(\dot c/c)^n v/c$ where $n$ is the rank
of the corresponding tensor in General relativity. Inertial forces
in these VSL theories may turn out to be a strong experimental
probe, and are currently under investigation.

%Finding the preferred frame outside cosmology is more arbitrary,
%and perhaps for this reason this theory has never been applied,
%say, to black holes.

\subsection{The gravitational equations}\label{cosmoeqs}
For the Friedmann metric, the equations in
Section~\ref{hardvsl} reduce to the familiar
\begin{eqnarray}
{\left({\dot a\over a}\right)}^2&=&{8\pi G\over 3}\rho -{Kc^2\over
a^2}
\label{fried1}\\
{\ddot a\over a}&=&-{4\pi G\over 3}{\left(\rho+3{p\over
c^2}\right)} \label{fried2}
\end{eqnarray}
where $\rho c^2$ and $p$ are the energy and pressure densities,
$K=0,\pm 1$ is the spatial ``curvature'', and a dot denotes a
derivative with respect to cosmological time. If the Universe is
flat ($K=0$) and radiation dominated ($p=\rho c^2/3$), we have as
usual $a\propto t^{1/2}$. As expected from minimal coupling, the
Friedmann equations remain valid, with $c$ being replaced by a
variable, in much the same way that Maxwell equations in media may
be obtained by simply replacing the dielectric constant of the
vacuum by that of the medium.

To follow up the discussion in Section~\ref{defvsl}, it is
possible to define units in which the speed of light is a
constant. Such a transformation is exhibited in \cite{barmag98},
and leads to the following equations for this theory:
\begin{eqnarray}
\left( {\frac{{\hat a}^{\prime }}{{\hat a}}}+{\frac{\epsilon ^{\prime }}%
\epsilon }\right) ^2 &=&{\frac{8\pi G}3}{\hat \rho }-{\frac{Kc_0^2}{{\hat a}%
^2},}  \label{fried1p} \\
{\frac{{\hat a}^{\prime \prime }}{{\hat a}}}+{\frac{\epsilon ^{,,}}\epsilon }%
-2{\left( \frac{\epsilon ^{\prime }}\epsilon \right) }^2 &=&-{\frac{4\pi G}3}%
({\hat \rho }+3{\hat p}/c_0^2),\label{fried2p}
\end{eqnarray}
where ${\hat a}$, ${\hat\rho}$ and ${\hat p}$ are expressed in the
new units, a prime denotes $d/d{\hat t}$ (where ${\hat t}$ is the
time in ``constant $c$' units), $\epsilon=c_0/c$, and ${\hat
K}=K=\{0,\pm 1\}$. For the dynamics of this theory choosing constant
$c$ units would be as silly as rephrasing Maxwell's equations in
units in which the speed of light remains constant even in dielectric
media. We hope that this mathematical illustration substantiates the
points made in Section~\ref{mean}.

\subsection{Violations of energy conservation}

Combining the two Friedmann equations (\ref{fried1}) and (\ref{fried2})
now leads to:
\begin{equation}\label{cons1}
\dot\rho+3{\dot a\over a}{\left(\rho+{p\over c^2}\right)}=
{3Kc^2\over 4\pi G a^2}{\dot c\over c}
\end{equation}
i.e. there is a source term in the energy conservation equation.
This turns out to be a general feature of this VSL theory. Energy
conservation derives, via Noether's theorem, from the invariance
of physics under time translations. The theory badly destroys the
latter, so it's not surprising that the former is also not true.

Another way to understand this phenomenon is to note that in
general relativity stress-energy conservation results directly
from Einstein's equations, as an integrability condition, via
Bianchi's identities. By tying our theory to a preferred frame,
violations of Bianchi identities must occur~\cite{bass}, and
furthermore the link between them and energy conservation is
broken. Hence we may expect violations of energy conservation and
these are proportional to gradients of $c$.

This effect brings out the most physical side of the issue of
mutability in physics~\cite{wheeler}. If we define mutability as a
lack of time translation invariance in physical laws, then what
might at first seem to be a metaphysical digression quickly
becomes a matter for theoretical physics. We find that lawlessness
carries with it shoddy accountancy. Not only do we lose the
concept of eternal law, but the book-keeping service provided by
energy conservation also goes out the window.

Creation of energy naturally leads to particle production~\cite{am}
and the VSL process of quantum particle creation was
further studied in \cite{partcreat}. Curiously, in the brane-world
realization of VSL (\cite{you1}) violations of energy
conservation are nothing but matter sticking to or falling off the
brane. The full theory (felt by the bulk) conserves energy
but the brane may fail to confine all matter fields, leading
to violations of energy conservation as perceived by observers
on the brane.

\subsection{The dynamics of the speed of light}

The issue remains as to what to make of $c$ itself. In the early
VSL work, the speed of light $c$, like the preferred frame
$u^\mu$, was taken as a given (in the same way that the dielectric
profile of a material is a given). Notice that in order to
explicitly break frame invariance, $u^\mu$ {\it should be preset}
and not be determined by the dynamics of the theory -- or else one
would only have broken frame invariance spontaneously. The same
cannot be said of the speed of light, which even in theories with
hard breaking of Lorentz invariance, may be seen either as a
pre-given law or as the result of a dynamical equation.

In the context of preset $c(t)$ two scenarios were considered:
phase transitions and the so-called Machian scenarios.
In phase transitions, the
speed of light varies abruptly at a critical temperature, as in
the models of \cite{moffat93,am}. This could be related to the
spontaneous breaking of local Lorentz symmetry \cite{moffat93}
(see Section~\ref{livsl} above). Later Barrow considered scenarios
in which the speed of light varies like a power of the expansion
factor $c\propto a^n$, the Machian scenarios. Taken at
face value such scenarios are inconsistent with experiment (see
\cite{vslsn} for an example of late-time constraints on $n$). Such
variations must therefore be confined to the very early universe
and so the $c$-function considered in \cite{ba,bm} should really
be understood as
\begin{equation}\label{cmachian}
c=c_0{\left(1+{\left( a\over a_0\right)}^n\right)}
\end{equation}
where $a_0$ is the scale at which VSL switches off (this point was
partly missed in~\cite{landau}) . The problem with putting in ``by
hand'' a function $c(t)$ is that the predictive power of the
theory is severely reduced (e.g.~\cite{mofwep}).

Later work~\cite{mofclay,vslsn}, however, endowed $c$ with a
dynamical equation (cf. Eqn.~\ref{dynabivsl}). For instance the
model described in Section~\ref{hardvsl} may be supplemented by an
equation of the form:
\begin{equation}
{\ddot \psi }+3{\frac{\dot a}a}{\dot \psi }=4\pi G\omega
f(\rho,p,K,\Lambda) \label{sor}
\end{equation}
where $\psi=log(c/c_0)$, and the function $f$ depends on the
model. If $f=\rho-p/c^2-(2Kc^2\omega/ a^2)$ one recovers the
Machian solution $c\propto a^n$ (with $n=\omega$), but this is
true for many other functions $f$\footnote{If we want to switch
off such variations at late times $\omega$ should be a function of
$\psi$, which goes to zero at a suitable value.}. For instance, in
\cite{vslsn} the choice $f=p/c^2$ was considered with the result:
\begin{equation}
n(t)\approx {\frac{\omega \rho _\gamma }{3(\rho _m+2\rho _\Lambda
)}}
\end{equation}
 The theory described in Section~\ref{livsl} also
predicts an equation of this form (cf. Eqn~\ref{dynacovvsl}), but
now it is the presence of the cosmological constant that drives
changes in $c$.

Whatever the source term chosen for this equation, Lambda has an
interesting effect upon the dynamics of $c$ -- the onset of Lambda
domination stabilizes $c$. This is a general feature of causal
varying constant theories; see \cite{bsbm} for a good discussion.
We shall return to this matter in Section~\ref{webbres},
in relation to the experimental status of these theories.

\subsection{Big Bang instabilities}
We will now be more quantitative regarding conditions for a
solution to the cosmological problems. We start with the horizon
problem. Let's take the phase transition scenario
\cite{moffat93,am} where at time $t_c$ the speed of light changes
from $c^-$ to $c^+$. Our past light cone intersects $t=t_c$ at a
sphere with comoving radius $r=c^+ (\eta_0-\eta_c)$, where
$\eta_0$ and $\eta_c$ are the conformal times now and at $t_c$.
The horizon size at $t_c$, on the other hand, has comoving radius
$r_h =c^-\eta_c$. If $c^-/c^+\gg\eta_0/\eta_c$, then $r\ll r_h$,
meaning that the whole observable Universe today has in fact
always been in causal contact. This requires
\begin{equation}\label{cond1}
\log_{10}{c^-\over c^+}\gg 32 -{1\over 2}\log_{10}z_{eq}+{1\over
2} \log_{10}{T^+_c\over T^+_P}
\end{equation}
where $z_{eq}$ is the redshift at matter radiation equality, and
$T^+_c$ and $T^+_P$ are the Universe and the Planck temperatures
after the phase transition. If $T^+_c\approx  T^+_P$ this implies
light travelling more than 32 orders of magnitude faster before
the phase transition. It is tempting, for symmetry reasons, simply
to postulate that $c^-=\infty$ but this is not strictly necessary.

Considering now the flatness problem, let $\rho_c$ be the critical
density of the Universe:
\begin{equation}
\rho_c={3\over8\pi G}{\left(\dot a\over a\right)}^2
\end{equation}
that is, the mass density corresponding to the flat model ($K=0$)
for a given value of $\dot a/a$. Let us quantify deviations from
flatness in terms of $\epsilon=\Omega-1$ with
$\Omega=\rho/\rho_c$. Using Eqns.(\ref{fried1}), (\ref{fried2})
and (\ref{cons1}) we arrive at:
\begin{equation}\label{epsiloneq}
\dot\epsilon=(1+\epsilon)\epsilon {\dot a\over a}
{\left(1+3w\right)}+2{\dot c\over c}\epsilon
\end{equation}
where $w=p/(\rho c^2)$ is the equation of state ($w=0,1/3$ for
matter/radiation). We conclude that in the standard Big Bang
theory $\epsilon$ grows like $a^2$ in the radiation era, and like
$a$ in the matter era, leading to a total growth by 32 orders of
magnitude since the Planck epoch. The observational fact that
$\epsilon$ can be at most of order 1 nowadays requires either that
$\epsilon=0$ strictly, or that an amazing fine tuning must have
existed in the initial conditions ($\epsilon<10^{-32}$ at
$t=t_P$). This is the flatness puzzle.

As Eqn.~(\ref{epsiloneq}) shows, a decreasing speed of light
($\dot c/c<0$) would drive $\epsilon$ to $0$, achieving the
required tuning. If the speed of light changes in a sharp phase
transition, with $|\dot c/c|\gg \dot a/a$, we find that a decrease
in $c$ by more than 32 orders or magnitude would suitably flatten
the Universe. But this should be obvious even before doing any
numerics,  from inspection of the non-conservation equation
Eq.~(\ref{cons1}). Indeed if $\rho$ is above its critical value
(as is the case for a closed Universe with $K=1$) then
Eq.~(\ref{cons1}) tells us that energy is destroyed. If
$\rho<\rho_c$ (as for an open model, for which $K=-1$) then energy
is produced. Either way the energy density is pushed towards the
critical value $\rho_c$. In contrast to the Big Bang model, during
a period with $\dot c/c<0$ only the flat, critical Universe is
stable. This is the VSL solution to the flatness problem.

VSL cosmology has had further success in fighting other problems
of Big Bang cosmology that are usually tackled by inflation. It
solves the entropy, isotropy, and homogeneity problems
\cite{moffat93,am}. It solves at least one version of the
cosmological constant problem~\cite{am} (see \cite{moflamb} for a
discussion of the quantum version of the lambda problem). It has a
quasi-flat and quasi-Lambda attractor~\cite{bm} (that is an
attractor with non-vanishing, but also non-dominating Lambda or
curvature).
More importantly, in some specific models VSL can lead to viable
structure formation scenarios~\cite{mofstruct,mofstruct1,pogo}, as shall be
discussed in Section~\ref{struc}.

\subsection{Further issues}

The cosmological implications of VSL have led to much further
work. The stability of the various solutions was demonstrated
in~\cite{dynamical,dynamical2} using dynamical systems methods
(see also \cite{chimento,gopa}). The role of Lorentz symmetry
breaking in the ability of these models to solve the cosmological
problems was discussed in \cite{avelvsl}. Also combinations of varying
$c$ and varying $G$ have been studied~\cite{bdicke,chak}.

Barrow \cite{unusual} has cast doubts on the ability of this model
to solve the isotropy problem, if the universe starts very
anisotropic. The issue is far from solved (the fact that the
universe is anisotropic does not preclude the existence of a
preferred frame, contrary to the assertion in \cite{unusual}). It
was also found that if $c$ falls off fast enough to solve the flatness
and horizon problems then the quantum wavelengths of massive
particle states and the radii of primordial black holes will grow
to exceed the scale of the particle horizon. However this
statement depends crucially on {\it how the Planck's constant
$\hbar$ scales with $c$ in VSL theories}. In \cite{am} one has
$\hbar\propto c$, but this need not be the case; see for instance
\cite{covvsl} (cf. Eqn.~\ref{alphan}).

The relation between VSL and the second law of thermodynamics was
investigated in \cite{diegp,youa,coule,unusual}. In particular
~\cite{diegp} found that if the second law of thermodynamics is to
be retained in open universes $c$ can only decrease, whereas in
flat and closed models it must stay constant. A similar discussion
in the context of black holes can be found in \cite{davis}. The
author would not be surprised if the second law of thermodynamics
were violated in VSL models with hard breaking of Lorentz
invariance. After all the first law is already violated.
However, a derivation from first principles remains elusive.
It is conceivable that the emergence of a macroscopic arrow
of time is made easier in VSL theories.

Finally the issue remains as to what the ultimate fate of the
universe will be. Taking as an example the model in \cite{vslsn},
we see that in addition to the solution found (which fits both the
supernovae and the changing alpha results),  there is a trivial
stable attractor: a non-VSL de Sitter universe. Even though this
solution is stable there is a minimal perturbation  (not
necessarily over a large volume) which takes it into the Big Bang
solution. This minimal perturbation should therefore be regarded
as a ``natural initial condition'' for the Big Bang solution. Such
a configuration --- the minimal point interpolating two stable
solutions -- is reminiscent of sphalarons in field theory.

Hence, in this model, a VSL Big Bang solution has a de Sitter
beginning, and given that it will end in $\Lambda $ domination,
the universe as a whole is de Sitter with sporadic Big Bang
``events''. At the end of a Big Bang phase $\Lambda $ dominates
and $c$ stops changing. Then a new fluctuation triggers another
VSL Big Bang event (\cite{qcosm} have suggested that such a
process is quantum). In every Big Bang cycle $c$ drops and $%
\Lambda $, measured  in fundamental units, increases. Its current
small value merely measures the number of cycles before ours,
starting from an arbitrarily small value. Therefore this scenario
not only explains why $\Lambda $ is only just beginning to
dominate, but also the smallness of $\Lambda $ in fundamental
units. Further implications for the anthropic principle may be
found in \cite{livio,anth}.
Needless to say these ``eschatological'' considerations are not
testable, and therefore belong to philosophy.

\section{The origin of cosmic structure}\label{struc}
The release of high-resolution CMB maps by the WMAP team
\cite{wmap} has opened the season for grand claims that
``inflation is proved''. Similar claims followed the first release
of the COBE-DMR maps \cite{cobe}, some ten years ago -- hardly an
exponent of signal to noise -- so perhaps a more sober assessment
of the implications of the latest observations is
required~\cite{vach}. What has actually been {\it observed}, and
in what way, if any, does it relate {\it directly} to inflation?

At a very low level of prejudice, WMAP revealed fluctuations that
look like the result of processing a Gaussian, nearly
scale-invariant (Harrison-Zeldovich) spectrum of initial
fluctuations through gravitational instability in a universe
filled with matter and radiation (and possibly a cosmological
constant). The amplitude of such a spectrum
is about $10^{-5}$ (see e.g.~\cite{pogo} for the relevant
definitions). The structure of Doppler peaks rules out a
significant contribution from {\it causal} defects (or more
generally from active incoherent sources \cite{afm}). Primordial
non-Gaussianity (with some unresolved hiccups \cite{nonG}), is
heavily constrained \cite{koma}.

Even though inflation produces this type of initial conditions
from ``first principles'', the recent observations are very far
from a direct observation of microphysical quantum fluctuations in
the inflaton field. In addition, one must recall that
Harrison-Zeldovich initial conditions were proposed
\cite{harr,zel} decades before inflation first saw the light of
day -- so their unambiguous association with inflation is denied
by history. There are also inflationary scenarios which produce
wildly different fluctuations \cite{linde,peebles}. Thus, it is
not clear how one may argue that the recent data proves inflation.
Rather, it is fair to say that the recent data strongly favors
Gaussian passive fluctuations with a Harrison-Zeldovich
spectrum~\footnote{More importantly, the data proves that gravity
is indeed the driving force of structure formation, and that the
theory of Jeans instability is correct.}.

However the recent data does impose a requirement for any would be
new scenario: alternatives, such as the ekpyrotik~\cite{ekp} or
VSL scenarios, should also produce a Harrison-Zeldovich spectrum
of passive Gaussian fluctuations. Of course  it is unlikely that
the CMB observations will be a more direct vindication of these
scenarios than of inflation. Nonetheless, this criterium acts as a
consistency condition for any alternative to inflation.

Even though the field is still in its infancy, we now summarize
the best attempts to derive the right structure formation
conditions in VSL scenarios.

\subsection{Quantum fluctuations in bimetric VSL}

Recently it was shown how the bimetric scalar-tensor model
described in Section~\ref{bivsl} could produce predictions in
agreement with the CMB spectrum observations, and therefore
provide an alternative scenario to the standard slow-roll
inflationary models~\cite{mofstruct,mofstruct1}. In these scenarios, depending
on the choice of frame, one may either have a fixed speed of light
and a dynamically-determined speed of gravitational disturbances
$v_g$, or a fixed speed of gravitational waves and a
dynamical speed of light. The former frame was chosen, and the
fluctuations in  the ``biscalar'' field $\phi$ were calculated in
a scenario where $v_g$ becomes vanishingly small in the very early
universe. When $v_g$ increases rapidly to its current value, the
effects of a potential $V(\phi)$ become important, and a quadratic
potential is introduced.

In this scenario it is found that scalar fluctuations have a
spectral index $n_s\approx 0.98$. There are also tensor
fluctuations with $n_t=-0.027$, and the ratio between the
amplitudes of tensor and scalar fluctuations satisfies $r\ge
0.014$. This is particularly important as upcoming polarization 
observations should constrain $r$. 
In contrast with inflation, this theory is not constrained by the
``consistency relation'' between $r$ and $n_s$ that is typical of
inflaton models. For a quadratic inflaton potential one would have
$r=7(1-n_s)/2$, and so the tensor modes are larger by a factor
of approximately $2$. Thus it could be that tensor modes could discriminate
between inflation and this VSL theory of structure formation.

Notice that the actual fluctuation mechanism in these VSL
scenarios is very similar to the inflationary one: vacuum quantum
fluctuations in the early universe, at a time when the comoving
horizons are shrinking.

A recalculation of fluctuations in this scenario, in the
frame in which the speed of light varies (and the speed of
gravity remains constant) was recently carried out~\cite{mofstruct1}.

\subsection{Thermal fluctuations}

More radically, it was suggested in \cite{pogo} that the cosmic
structure could have a thermal origin. This possibility was first
advanced by Peebles, (\cite{peebook}, pp.~371-373) who pointed out
that if the Universe was in thermal equilibrium on the comoving
scale of 10 Mpc when its temperature was $T=10^{11}$~Gev, then the
observed value of $\sigma_{10}$ could be explained. The question
arises as to how such a large scale could be in thermal
equilibrium -- and thus in causal contact -- at such an early
time. VSL provides an explanation, as shown in \cite{pogo}.

Thermal fluctuations are Gaussian to a very good approximation;
however they are white-noise ($n_s=0$), rather than
scale-invariant ($n_s=1$). More precisely, the power spectrum of
thermal fluctuations is:
\begin{equation}
\label{amplT}
P(k)={\langle {|\delta_k|^2} \rangle} \approx {T^2\rho'\over \rho^2} k^0
\end{equation}
(where we have ignored factors of order 1). This result
depends only on the form of the partition function, and is true even if
the dispersion relations are deformed (see Section~\ref{colvsl}).
The white noise nature of the spectrum only makes use of the fact
that energy is an extensive quantity (i.e. proportional to the
volume).

This result only applies to modes which are in causal contact,
i.e. sub-horizon modes. As the modes leave the horizon they
freeze in. Thus the spectrum left outside the horizon could be
scale-invariant, depending on the timing of horizon leaving. It
was shown in \cite{pogo} that this dynamics depends crucially on
the (deformed) Stefan-Boltzmann law $\rho\propto T^\gamma$
resulting from deformed dispersion relations. If the universe is
cooling this requires that $\gamma<1$ and it was shown that this
cannot be achieved for generic particle gases. Therefore the most
promising scenarios are those in which the universe gets hot in
time (``warming universes'') while density fluctuations are being
produced.

Surprisingly, one such scenario is a bouncing universe:
a closed universe that goes through a series of cycles starting
with a Big Bang and expansion, followed by re-contraction and a
Big Crunch. It can be shown~\cite{pogo} that even if such a universe
is filled with undeformed radiation ($\gamma=4$), modes leave the horizon
as the universe contracts, and the frozen-in thermal fluctuations are
indeed scale-invariant. This is a general result (at least if one
ignores the issue of mode-matching at the bounce), and dispenses with
fine-tuning. However in general the amplitude of such a spectrum
is of order 1 (rather than the observed $10^{-5}$), so such a universe
is grossly inhomogeneous. It is at this stage that introducing a variation
in the speed of light {\it at the bounce} introduces the right tuning.

Let us illustrate the argument using the $10$ Mpc comoving scale
as the normalization point. The correct normalization can be obtained
if the 10 Mpc comoving scale leaves the horizon in the contracting phase
when the universe is at $10^{11}$ Gev~\cite{peebles}. However, if
no constants vary at the bounce,  this scale will leave the
horizon much earlier  in the contracting phase, when the universe is
colder and therefore the fluctuations much larger. The only way
to fix the normalization is to allow for a change in the values of the
constants at the bounce. If one assumes that the relation
between time and temperature is symmetrical around the bounce, but
that the speed of light in the previous cycle $c_-$ is much larger
than nowadays ($c_+$), then the $10$ Mpc comoving scale would leave the
horizon at a higher temperature. More concretely one finds:
\begin{equation}\label{c-c+}
{c_-\over c_+}
\approx 10^{21}
\end{equation}
in order to produce the appropriate normalization.
Note that unlike other VSL arguments, which lead to lower bounds on
$c_-/c_+$, this argument leads to an identity: the spectrum amplitude
results directly from a given value of $c_-/c_+$.

The issue of tensor modes in these scenarios is far more unsure.

\subsection{Other scenarios}
There are other successful scenarios.  For instance \cite{defvsl}
have shown how, in bimetric VSL, topological
defects could produce a Harrison-Zeldovich spectrum of primordial
density perturbations. This would happen while the speed
associated with the defect-producing scalar field was much larger
than the speed of gravity and all standard model particles. Such a
model would exactly mimic the standard predictions of inflationary
models (apart from leaving traces of non-Gaussianity). Therefore
it could even be that the current observations are the result of
``acausal'' defects.

Fluctuations in the scenarios described in Section~\ref{hardvsl} have also
been studied using the gauge invariant formalism (see also \cite{gauge}
for related work).
It was shown \cite{am,bdicke} that the
comoving density contrast $\Delta $ and gauge-invariant velocity $v$ are
subject to the equations:
\begin{eqnarray}
\Delta ^{\prime }-{\left( 3(\gamma -1){\frac{a^{\prime }}a}+{\frac{c^{\prime
}}c}\right) }\Delta &=&-\gamma kv-2{\frac{a^{\prime }}a}(\gamma -1)\Pi _T
\label{delcdotm} \\
v^{\prime }+{\left( {\frac{a^{\prime }}a}-2{\frac{c^{\prime }}c}\right) }v
&=&{\left( {\frac{c_s^2k}\gamma }-{\frac 3{2k}}{\frac{a^{\prime }}a}{\left( {%
\frac{a^{\prime }}a}+{\frac{c^{\prime }}c}\right) }\right) }\Delta  \nonumber
\\
+{\frac{kc^2(\gamma -1)}\gamma }\Gamma - &&\ kc(\gamma -1)\left( \frac
2{3\gamma }+\frac 3{k^2c^2}\left( \frac{a^{\prime }}a\right) ^2\right) \Pi _T
\label{vcdotm}
\end{eqnarray}
where $k$ is the comoving wave vector of the fluctuations, the
dash denotes derivatives with respect to conformal time,
$p=(\gamma-1)\rho$, $\Gamma $ is the entropy production rate, $\Pi
_T$ the anisotropic stress, and the speed of sound $c_s$ is given
by
\begin{equation}
c_s^2={\frac{p^{\prime }}{\rho ^{\prime }}}=(\gamma -1)c^2{\left( 1-{\frac
2{3\gamma }\ \frac{c^{\prime }}c}{\frac a{a^{\prime }}}\right) }  \label{cs}
\end{equation}
The free, radiation dominated solution for the Machian scenario
(cf. Eqn.~\ref{cmachian}) is
\begin{equation}
\Delta =A\eta ^{2(n+1)}+B\eta ^{n-1}
\end{equation}
where $A$ and $B$ are constants in time. For a constant $c$ ($n=0$) this
reduces to the usual $\Delta \propto \eta ^2$ growing mode, and $\Delta
\propto 1/\eta $ decaying mode. Carefully designing $c(t)$ could again
convert a white noise thermal spectrum into a scale invariant spectrum.

\section{VSL and quantum gravity}\label{qg}
It is likely that varying speed of light theories will play an
important part in the quest for a theory of quantum gravity~\cite{leereview},
although their exact role is at the moment unclear. A simple
argument underpins this assertion. The combination of gravity
($G$), the quantum ($\hbar$) and relativity ($c$) gives rise to
the Planck length, $ l_P = \sqrt{\hbar G / c^3 } $, the Planck
time $t_P=l_P/c$, and the Planck energy $E_P= h/t_P$. These scales
mark thresholds beyond which the classical description of
space-time breaks down and qualitatively new phenomena are
expected to appear. No one knows what these new phenomena might
be, but both loop quantum gravity \cite{rovelli,carlip} and string
theory \cite{pol,stringref} are expected to make clear predictions
about them once suitably matured.

However, whatever quantum gravity  may turn out to be, it is
expected to agree with special relativity when the gravitational
field is weak or absent, and for all experiments probing the
nature of space-time at energy scales much smaller than $E_P$
(or length/time scales larger than $l_P$ or $t_P$).
This immediately gives rise  to a simple question:  {\it in whose
reference frame are $l_P$, $t_P$ and $E_P$ the thresholds for new
phenomena?} For suppose that there is a physical length scale
which measures the size of spatial structures in quantum
space-times, such as the discrete area and volume predicted by
loop quantum gravity. Then if this scale is $l_P$ in one inertial
reference frame, special relativity suggests it may be different
in another observer's frame: a straightforward implication of
Lorentz-Fitzgerald contraction.

There are several different answers to this question, the most
obvious being that Lorentz invariance (both global and local) may
only be an approximate symmetry, which is broken at the Planck
scale. One may then correct the Lorentz transformations so as to
leave the Planck scale invariant. Corrections of Lorentz symmetry
require violations of one or both of its underlying postulates:
the relativity of motion and the constancy of the speed of light. Thus VSL
sneaks into quantum gravity considerations.

Other solutions to the problem are possible (e.g. \cite{rovellidsr}).
For instance, one feasible response
is that the question does not make sense. In $S$ matrix type
theories (such as string theory), reality is made up of scattering
experiments, for which a preferred frame is always present: the
center of mass frame. The existence of this frame does not violate
special relativity, and establishes an invariant division between
the realm of classical and quantum gravity. It is, however, extremely
unsatisfactory to reduce reality to scattering experiments. The world
is not a huge accelerator.

In this section we
 review recent work where a varying speed of light is employed
to introduce an invariant energy and/or length scale. In these
theories, if a phenomenon is ``classical'' in one frame, it will
never appear as ``quantum'' in any other frame. The division between
classical and quantum gravity is thus invariant. A number of interesting
physical and mathematical results follow from this requirement.
The subject has been partly reviewed in \cite{camreview},
and we will avoid overlap.

\subsection{Non-linear realizations of the Lorentz group}
The starting point for much of the work in this field is the
possibility of departures from the usual dispersion relations
$E^2-p^2=m^2$. This is observationally motivated by anomalies in
ultra high-energy cosmic ray protons~\cite{review,crexp,cosmicray},
as well as (possibly)  Tev photons~\cite{gammaexp}. Such
dispersion relations also lead to an
energy dependent speed of light, observable in planned gamma ray
observations~\cite{amel,amel1,liouv}.

An example of a deformation
was given in Eqn.~(\ref{dispcam}), but more generally one may
consider deformations of the form:
\begin{equation}\label{desp}
  E^2f_1^2(E;\lambda)-p^2f_2^2(E;\lambda)=m^2
\end{equation}
where $f_1$ and $f_2$ are phenomenological functions. If one
insists~\cite{amel,amel1,liouv} that Lorentz
transformations remain linear, then Eqn.~(\ref{desp}) can only be
true in one frame; thus the principle of relativity is violated.
However this need not be true~\cite{amelstat,gli,leejoao,leejoao1}, if one
allows for a non-linear form for the Lorentz transformations. If
the ordinary Lorentz generators act as
  \f L_{ab} = p_a {\partial \over \partial p^b} -
 p_b {\partial \over \partial p^a}
\ff then we consider the modified algebra
\begin{equation}\label{U}
K^i = U^{-1} [p_0] L_0^{\ i} U [p_0 ]
\end{equation}
with \f U \circ (E, {\bf p})=(Ef_1,{\bf p}f_2). \label{udef} \ff
The form of the Lorentz algebra (its commutators)
is still the same, but the group
realization is now non-linear. Following these modified Lorentz
transformations  the dispersion relations (\ref{desp}) are true in
any frame, so that the principle of relativity has been
incorporated into the theory (see also \cite{lehn}).

A concrete example \cite{leejoao} results from
\begin{equation}\label{U1} U [p_0]\equiv \exp(\lambda p_0 D)
\end{equation}
where $D=p_a{\partial\over \partial p_a}$ is a dilatation and
$\lambda$ a given length scale. It is associated with the
dispersion relations \f\label{invariant} {\eta^{ab}p_a p_b \over
(1-l_P p_0 )^2} =m^2 \label{inv} \ff and results in the
transformation laws
 \bea \label{fltransp}
p_0'&=&{\gamma\left(p_0- vp_z \right)
\over 1+l_P (\gamma -1) p_0  -l_P\gamma v p_z }\\
p_z'&=&{\gamma\left(p_z- v p_0\right)
\over 1+l_P (\gamma -1) p_0  -l_P\gamma v p_z }\\
p_x'&=&{p_x
\over 1+l_P (\gamma -1) p_0  -l_P\gamma v p_z }\\
p_y'&=&{p_y \over 1+l_P (\gamma -1) p_0  -l_P\gamma v p_z } \eea
which reduce to the usual transformations for small $|p_\mu|$.
This choice for $U$ is quite distinct from that in \cite{ameldsr},
and an appraisal and comparison may be found in
\cite{giogen,giocomp,leejoao1}. Further examples may be
found in~\cite{heud}.

Since the structure of the algebra remains the same
\cite{ahl}, and indeed the variables $U(p_a)$ transform
linearly, claims have been made that the theory is physically
trivial~\cite{ahl}. This is clearly not the case; for
instance the redshift formula in the theory proposed in
\cite{leejoao} is \be {\Delta E\over E}=\Delta \phi{\left(
1-l_PE\right)} \ee with clear implications for the Pound-Rebbka
experiment~\cite{will}. Mathematical triviality by no means
implies physical equivalence, and one may argue that it is in fact an
asset. If new physical results may be derived from a different
representation of the same old group then such a formulation may be
preferable. This matter was further discussed in \cite{gonzo,gonzo1}.

A related argument was put forward in~\cite{toller1,Grum},
where gravity was examined. It was found that either the metric is
non-invariant (as in \cite{am}, and also in~\cite{mignemi2}), or an
equivalence with the undeformed metric is established. This clearly depends
on how to introduce gravity into the theory; see
for instance~\cite{grav,twod}.

\subsection{Physical properties and the choice of $U$}\label{physdsr}
The $U$ map can be chosen so as to implement various properties
required from a phenomenological theory of quantum gravity, such
as an invariant energy scale and a maximal momentum. These, as we
shall see, require the existence of an energy dependent speed of
light. Thus the theories usually referred to as DSR (doubly
special relativity~\cite{ameldsr}) form a subclass of VSL
theories.

Before we consider the physical constraints upon $U$ we must first
note that there are a number of consistency conditions which
$U$ must satisfy. For the action of the Lorentz group to be
modified according to (\ref{U}), $U$ must be invertible. Generally
this restricts the physical momentum space. Typical examples of
restriction are $E < E_{Planck}$ and/or $|{\bf p }| < E_{Planck}$.
A further condition is that the image of $U$ must include the
range $[0,\infty]$ for both energy and momentum. This is because
the ordinary Lorentz boosts $L$ span this interval, and so
$U^{-1}LU$ would not always exist otherwise. If this condition is
not satisfied the group property of the modified Lorentz action is
destroyed. If for instance $Ef_1$ does not span $[0,\infty]$ then
there is a limiting $\gamma$ factor for each energy, a feature
which not only destroys the group property but also selects a
preferred frame, thereby violating the principle of relativity.
(That the group property is satisfied by Eqns.~(\ref{fltransp})
has been explicitly verified by \cite{bruno}).

Bearing this in mind we may now lay down the conditions upon $U$
for an invariant energy scale (and so an invariant separation
between classical and quantum space-time). Given (\ref{U}) we know
that the invariants of the new theory are the inverse images via
$U$ of the invariants of standard special relativity. But the only
invariant energies in linear relativity are zero and infinite. The
case of zero energy may be easily ruled out~\cite{djj}. Hence the
condition for a non-linear theory to display an invariant Planck
energy is
\begin{equation}\label{uep}
  U(E_P)=E_pf_1(E_P)=\infty
\end{equation}
that is, $U$ should be singular at $E_P$.

The conclusion is that, under this condition, it is possible to
have complete relativity of inertial frames, and at the same time
have all observers agree that the scale at which a transition from
classical to quantum spacetime occurs is the Planck scale, which
is the same in every reference frame. Since the non-linear
transformations reduce to the familiar and well tested actions of
Lorentz boosts at large distances and low energy scales, no
obvious conflict with experiment arises.

The connection with VSL arises because unless $f_1=f_2$  we obtain
a theory displaying a frequency dependent speed of light. Defining
$f_3=f_2/f_1$ we have
\begin{equation}\label{speedc}
c={dE\over dp}={f_3\over 1-{Ef_3'\over f_3}}
\end{equation}
which is usually the group velocity of light. This definition has
been contested~\cite{remb,Kos,mignemi} on a variety of grounds. If
one takes $v=dx/dt$ as the primary definition of velocity
(relevant in gamma-ray timing experiments \cite{amel}), one needs
to rediscover position space from the usual momentum space
formulation to find the correct answer~\cite{djj}. In some
theories~\cite{djj,Kos}, this results in
\begin{equation}
c=\frac{E}{p}=f_3.
\end{equation}
Different answers may also be obtained if one is prepared to
accept a non-commutative geometry~\cite{tezuka,velnonc,velnonc1}.
In any case all formulations agree that if $f_1\ne f_2$ there is
an energy dependent speed of light.

If we now impose that there must be a maximal momentum (another
desirable property in quantum gravity theories) then we must
require that $Ef_1/f_2=E/f_3$ has a maximum. Hence  the conditions
for a varying speed of light and for the existence of a maximum
momentum are related. One may show that the existence of a maximum
momentum implies that the speed of light must diverge at some
energy~\cite{leejoao1}.

\subsection{Threshold and gamma-ray anomalies and other experimental tests}
\label{threshdsr}
 Besides its motivation as a phenomenological
description of quantum gravity, non-linear relativity has gained
respectability as a possible solution to the puzzle of threshold
anomalies\cite{amelstat,gli} (see also
\cite{cosmicray,amel,amel1,liouv,gamb,alf,ad,ncvsl} for other
experimental implications).

\begin{figure}\label{fig3}
%\centerline{\includegraphics{uhecr.ps}}
\begin{center}
\psfig{file=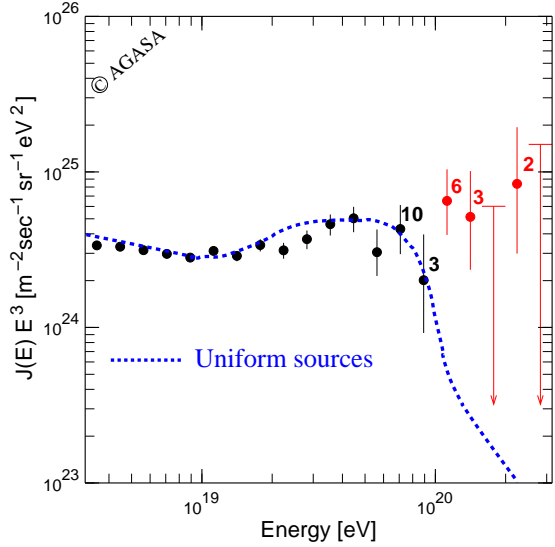,width=8cm} \caption{The flux of cosmic rays
at high energies. The dashed line illustrates the GZK cut-off.
(I thank Masahiro Takeda for permission to use this Figure.)}
\end{center}
\end{figure}

Ultra high energy cosmic rays (UHECRs) are rare showers derived
from a primary cosmic ray, probably a proton, with energy above
$10^{11}$ Gev. At these energies there are no known cosmic ray
sources within our own galaxy, so it's expected that in their
travels, the extra-galactic UHECRs interact with the cosmic
microwave background (CMB). These interactions should impose a
hard cut-off above $E_{th0}\approx 10^{11}$ Gev, the energy at
which it becomes kinematically possible to produce a pion. This is
the so-called Greisen-Zatsepin-Kuzmin (GZK) cut-off; however
UHECRs have been observed beyond the threshold \cite{review,crexp}
(see Fig.~\ref{fig3}).  A similar threshold anomaly results from
the observation of high energy gamma rays above 10
Tev~\cite{gammaexp}, but in this case it's far less obvious that
there is indeed an observational crisis.

We now show how non-linear relativity modifies these thresholds.
We present a modified argument proving that threshold anomalies do
not depend on the complex issue of how to formulate energy and
momentum conservation in non-linear relativity. For pedagogical
reasons we take as an example gamma rays, for which one expects a
cut-off due to interactions with the infra-red background. The
cut-off energy corresponds to the kinematical condition for
production of an electron-positron pair out of a gamma ray and an
infra-red photon \cite{gammaexp}. For a threshold reaction, the
electron and positron have no momentum in the center of mass
frame. Furthermore,  in this frame the gamma ray and the infra-red
photon have the same energy. (Undeformed) energy conservation then
implies that both photons have an energy equal to the rest energy
of an electron $m_e$. We can draw these conclusions since $m_e\ll
E_P$, so that any corrections imposed by our theory are
negligible.

We then need to perform a boost transformation from the center of
mass frame to the cosmological frame. This can be pinned down by
the condition that one of the photons be redshifted to the
infra-red background energy. Since in this process all energies
involved are again sub-Planckian we can use plain special
relativistic formulae to conclude that $E_{IR}=(1-v)\gamma m_e$,
and since $\gamma\gg 1$ (implying $1-v\approx 1/(2\gamma^2)$) we
have:
\begin{equation}
\gamma={m_e\over 2E_{IR}}
\end{equation}
The same boost transformation blueshifts the other photon to our
predicted value for the gamma ray threshold energy. This
operation, however, has to be performed with the corrected
non-linear boost. The uncorrected threshold energy is
\begin{equation}
E_{th0}=\gamma(1+v)m_e\approx 2\gamma m_e={m_e^2\over E_{IR}}
\end{equation}
This now becomes:
\begin{equation}\label{thrgamma}
E_{th}=U^{-1}(E_{th0}) \label{problem2}
\end{equation}
since the non-linear boost derives from $U(E_{th})=\gamma(1+v)U(m_e)$, and
we have $U(m_e)\approx m_e$. We thus arrive at (\ref{thrgamma}) without
using correction to (linear) energy conservation, using only one
non-linear boost.

For UHECRs one considers the interaction of protons and CMB
photons, and the threshold for pion production
(e.g.~\cite{leejoao1}). An identical argument leads to the
corrected threshold:
\begin{equation}
E_{th}=U^{-1}(E_{th0})=U^{-1}{\left({(m_p+m_\pi)^2-m_p^2\over
E_{CMB}}\right)}
\end{equation}
where $m_p$ and $m_\pi$ are the proton and pion rest masses, and
$E_{CMB}$ is the photon CMB energy in the cosmological frame. It
was shown in \cite{leejoao1} that it is possible to resolve the
UHECR anomaly, while preserving an invariant energy scale of the
order of the Planck energy. One may choose, for instance,
\begin{equation}
f_1={1\over (1+\lambda_1 E)(1-\lambda E )}
\end{equation}
with $\lambda_1^{-1}\approx 10^{11}Gev$ and $E_P=\lambda^{-1}$
(which may be of order $10^{19}$ Gev).

Another solution  is to allow for a non-universal
$U$~\cite{giosolution}. Different particles could then have a
different $U$, or $U$ could depend on the rest mass $m$ of the
particle on which it acts. One can then use the proton mass as an
automatic extra scale in the problem, ensuring that the Planck
scale is invariant and that the UHECRs threshold is raised.

A large number of variations around this theme have been put
forward, e.g.~\cite{Toller2,lehn}. On a very different front,
spontaneous symmetry breaking of Lorentz
invariance~\cite{gzkmoffat} has also been used as a possible
explanation for threshold anomalies.

Other implications of non-linear relativity for other areas of
high energy astrophysics have been studied. Processes such as
vacuum Cerenkov radiation ($e^-\rightarrow e^-\gamma$), photon
decay $\gamma\rightarrow e^+e^-$, and other exotic processes are
kinematically possible in some of these theories, leading to
constraints upon their parameter space \cite{jacob,limits}.
Synchrotron emission from the Crab nebula was used in \cite{crab}
to  further tighten these constraints, but this work has been
criticized for being model dependent (in the sense that it depends
on more than kinematical arguments) \cite{amecritic,replyac}.

Finally the authors in~\cite{myers} have used effective field
theory to to find dimension 5 operators that do not mix with
dimensions 3 and 4 and lead to cubic modifications of dispersion
relations for scalars, fermions, and vector particles. Clock
comparison experiments bound these operators at $10^{-5}/E_P$.

\subsection{Theoretical developments}\label{theoryqg}
The outline of non-linear relativity given above leaves many
questions unanswered, and considerable work is currently in
progress.

Foremost, there is the problem of how to deal with composite
systems. With the loss of linearity, the kinematic relations valid
for single particles need not be true for composite systems. This
is a desirable feature: {\it non-linearity builds into the theory
the concept of elementary particle}, clearly differentiating
between fundamental particles and composites.

The matter appears at first when defining a covariant law of
addition of energy and momentum. The most straightforward
definition (chosen in \cite{judvisser}) is
\be
p_a^{(12)}=U^{-1}(U(p^{(1)}_a)+U(p^{(2)}_a))
\ee 
but it quickly leads to
inconsistencies, e.g. it  implies that a set of particles
satisfying $E\ll E_P$ can never have a collective energy larger
than $E_p$. This is blatantly in contradiction with observation. A
possible solution is to note that the $U$ used for a system of,
say, two particles need not be the same as that used for single
particles, with
$p_a^{(12)}=U^{-1}_{2}(U(p^{(1)}_a)+U(p^{(2)}_a))$. A possible
choice for a system of $n$ particles is then
$U_n=U[p_0;\lambda/n]$, that is a system of $n$ elementary
particles should satisfy kinematical relations obtained from a map
for which the Planck energy $E_P=\lambda^{-1}$ is replaced by
$nE_P$.

Thus:
\begin{eqnarray}
p^{(12)}&\equiv &p^{(1)}\oplus p^{(2)}\nonumber\\
&=&U^{-1}\left[p_0;\lambda/2 \right]((U[p_0;\lambda
](p^{(1)})+U[p_0; \lambda](p^{(2)}))
\end{eqnarray}
This defines a new, generally nonlinear, composition law for
energy and momenta, which we denote by $\oplus$ to indicate that
it is not ordinary addition. In general
\begin{equation}\label{add2}
p^{(1...n)}=U^{-1}[p_0;\lambda /n ](U[p_0; \lambda
](p_1)+...+U[p_0;\lambda ](p_n))
\end{equation}
With this definition a system of $n$ particles satisfies a system
of transformations obtained from $U[p_0;\lambda/n]$ via (\ref{U}),
equivalent to the usual ones but replacing $\lambda $ with
$\lambda /n$. As a result, the collective momentum
$P^{(N)}=p^{(1...n)}$ satisfies deformed dispersion relations with
$\lambda $ replaced by $\lambda /n$. This can never lead to
inconsistencies, because if all $n$ particles of a system have
sub-Planckian energies then the total will still be sub-Planckian,
in the sense that $E_{tot}\ll nE_P$. This addition law is generally
commutative but non-associative, a feature
expected of any addition law incorporating
the concept of elementary particle.

There are other solutions to this problem. The embedding proposed
in~\cite{desitter} amounts to a choice of $U_2$ which depends on
the momentum of each individual particle according to a well
defined formula.

Once addition rules are defined, the law of energy and momentum
conservation  is straightforward to implement; with the obvious
exception of processes in which the number of particles changes.
For an example of energy conservation in a process where different
particles satisfy different dispersion relations see~\cite{neut}.

Another theoretical development concerns the position space picture of these
theories (which are more usually
constructed in momentum space). With loss of
linearity, duals no longer mimic one another, that is, vectors no
longer transform according to the inverse (linear) transformation
of co-vectors. A number of solutions may be found, either
involving \cite{velnonc} or avoiding non-commutative
geometry~\cite{djj}. The possible relation to quantum groups has
long been known
\cite{kowal1,piotr,Luk,lukie,kappa,noncom,gli,majid,lukie,granik,ball}
(see also \cite{Gob}).

It is obvious that one may recover linearity by embedding the
theory into a higher number of dimensions. This approach is very
elegant, and has been implemented in \cite{desitter} (where an
identification with deSitter space is accomplished) and in
\cite{embed}. An alternative way to linearize the theory is to
introduce a modified boost
parameter~\cite{judvisser,giogen,giocomp}.

A considerable amount of work has also gone into formulating field
theory. Canonical quantization, and the set up of the Fock space
have been examined \cite{leejoao1,dadic}. While scalar fields are
straightforward to implement \cite{leejoao1,djj}, fermions are far
more complicated \cite{alf1,giospin,fermi}. Nonetheless,
\cite{Chak} showed how to produce non-linear realizations of the
Poincar\'e group for arbitrary mass and spin. A supersymmetric
extension of the Poincar\'e algebra was also studied.

Another field theory ramification resulted from the realization
that neutrino flavour states do not satisfy standard dispersion
relations and therefore transform according to a non-linear
realization of the Lorentz group~\cite{neut}. This has dramatic implications
for the endpoint of $\beta$ decay; for instance one may accommodate
a fit with $m_\nu^2<0$ (as suggested by experiment)
without introducing acausal behaviour. In general it is found
that $f_1^2$ does not need to be positive and so  $m^2<0$
and tachyonic acausal behaviour cannot be identified. This feature
was used to remove the tachyonic ground state of bosonic string theory
in \cite{ljstring}.

But perhaps the most tantalizing theoretical problem faced
by this approach is how to do general
relativity based on non-linear relativity~\cite{mignemi2,grav}.  If
one builds position space so that the space-time transformations
are still linear (albeit energy dependent), a metric can still be
defined, even though the metric tensor is now energy
dependent~\cite{djj}. In some sense, the metric ``runs'' with the
energy, and general relativity can be set up in a straightforward
fashion~\cite{grav}. If, on the other hand, at high energy (or at
small distances) the concept of metric simply disintegrates, the
problem becomes potentially much more complicated.

%So much so that we will reserve discussion of this project for the
%last section of this review, where work in progress will be
%described.

\section{``Black'' holes and other compact objects}\label{bh}

A major issue under investigation is the fate of black holes in
the various VSL theories. This was first studied~\cite{vslbh} in
the context of the VSL theories presented in Section~\ref{livsl}.
The general aspect of the solutions is very similar to the
solutions previously found by Brans and Dicke~\cite{bdick}: \bea
ds^2&=&-F^{2\over \lambda}d\xi^2 + {\left(1+{\rho_0\over
\rho}\right)}^4 F^{2(\lambda-C-1)\over \lambda}
(d\rho^2 +\rho^2 d\Omega^2)\label{bd1}\\
c&=&c_0F^{C\over a \lambda}\label{bd2}
\eea
with
\bea
F&=&{1-\rho_0/\rho\over 1+\rho_0/\rho}\\
\lambda^2&=&(C+1)^2-C(1-\kappa C/(2a^2))\label{lamb} \eea (see
\cite{bdick} for an explanation of the coordinates used,
and~\cite{vslbh} for a relation between the various constants in
this solution and the parameters of the theory). In a limiting
case of the parameters of the theory, one may also recover the
Schwarzschild solution, in which case $c$ varies in space
according to: \be\label{cschw} c=c_\infty{\left(1-{2Gm\over
c_\infty^2 r}\right)}^{b-a\over 2\kappa} \ee We see that the speed
of light either goes to zero or to infinity at the horizon (depending
on the couplings' signs). This property can be generally proved, and a
simple condition for $c=0$ at the
horizon was identified in \cite{vslbh}.

This result suggests that the singularity at $r=0$ (and in some
cases at the horizon) is physically inaccessible, not just in the
sense that information cannot flow from it into the asymptotically
flat region, but also in the sense that no observer starting from
the asymptotically flat region can actually reach it. The
singularity lies in a disconnected piece of the manifold, which
should simply be excised as unphysical. It is tempting to
conjecture that all singularities are subject to the same
constraint, in which case VSL seems to have eliminated the
singularity problem, by means of a stronger version of the cosmic
censorship principle.  A similar argument regarding the
cosmological singularity in Brans-Dicke theories was put forward
in \cite{quiroz}. However these remarks still belong to the realm
of conjecture.

More generally, the removal of singularities in VSL may be related
to the presence of a maximal acceleration~\cite{maxacc,maxacc1} in
some of these theories. An interesting condensed matter analogy
may be found in \cite{volo}, where quantum liquids are used to
simulate the behavior of the quantum vacuum in the presence of the
event horizon. It is found that in most cases the quantum vacuum
resists the formation of the horizon.

Since the speed of light vanishes at the Schwarzschild radius, it
is not surprising that stellar collapse~\cite{vslbh} and the
properties of compact objects~\cite{neutstar} are ultimately very
different in these theories. The Oppenheimer-Snyder solution, in
which a spherical dust ball collapses, was found in \cite{vslbh}.
As the surface of the star approaches its Schwarzschild radius,
all processes freeze-out. We are left with a Schwarzschild-sized
remnant. Neutron stars were also extensively studied in
\cite{neutstar}. For certain choices of parameters VSL neutron
stars are much smaller than those in GR, and the dependence of their
mass on the strength of the coupling between $\psi$ and matter is
extremely strong. Thus the existence of neutron stars was used to
place constraints on the theory's parameters (even though the
analysis is very sensitive to the equation of state used).

Black hole thermodynamics in VSL theories has also been studied
under simplified assumptions, namely that the usual area formulae
remain valid. Assuming a fixed black hole mass and the standard
form of the Bekenstein-Hawking entropy, \cite{davis} have argued
that the laws of black hole thermodynamics disfavor models in
which the fundamental electric charge $e$ changes. Indeed if the
black hole entropy is still given by \be S={k_b\pi G\over \hbar
c}{\left(M+{\sqrt{M^2 + Q^2/G}}\right)}^2\ee and if $\hbar$, $G$
and $M$ remain constant, there is a clear difference between
varying-$c$ and varying-$e$ theories. With similar assumptions,
\cite{carlipbh} then showed that severe constraints apply to VSL.
This issue is related to the fate of the second law of
thermodynamics as discussed in \cite{diegp}, and work in this area
may be criticized along the same lines used above. Foremost, it is
probable that particle production (and so Hawking radiation) would
have to be radically modified (as suggested in \cite{covvsl}). A
more complete study of this issue in dilaton
theories may be found in~\cite{fairb}; see also~\cite{vagenas}
for a 2D stringy black hole.

For further work on black holes and varying constants
see~\cite{bhiran,bhfar}.  It should be obvious that deformed
dispersion relations (as those encoded by non-linear relativity)
will translate into a non-thermal spectrum for the Hawking
radiation. Some preliminary work on this topic may be found in
\cite{grav,casadio} (see also \cite{cas1,cas2}).

\section{The observational status of VSL}\label{obstat}

In the middle of the current observational revolution in
cosmology, it's easy to forget that some nasty surprises have also
fallen from the sky. Examples include claims for cosmic
acceleration~\cite{super,super1,super2,super3}, or the mounting
evidence~\cite{webb,murphy} for a redshift dependence in the fine
structure constant $\alpha$. Cosmologists can no longer, as in
Dirac's quote opening this paper, make ``any assumptions that they
fancy''; instead it appears that they must grapple with the issue
of selecting which observations to take seriously. Most of what
passes for observation in cosmology is plagued by systematic
errors. Some of these ``facts'' could evaporate like fog should a
new technological revolution come on line unexpectedly.

It is nonetheless interesting that several observational puzzles
can be solved by VSL. With appropriate supplementary observations,
the redshift dependence in $\alpha$ could be seen as the result of
a varying $c$. Another puzzle, already studied in
Section~\ref{threshdsr}, was the observation of rare {\it very}
high energy cosmic rays, in conflict with standard kinematic
calculations based on special relativity, which predict a cut-off
well below the observed energies. This could represent the first
experimental mishap of special relativity, and evidence for some
VSL theories. Finally, even the accelerating universe may be part
of a varying $c$ picture of the world.

How can this meager evidence be extended? Unfortunately there are
two obstacles to the observation of a varying speed of light. The
first relates to the discussion in Section~\ref{mean} and affects
those aspects of VSL which are not dimensionless. It is easy to
place oneself in a no-win situation (e.g. by defining units in
which $c$ is a constant), but as explained in Section~\ref{mean}
the impasse may be solved by testing the dynamics of the theory.
The issue of testability is more direct regarding the
dimensionless aspects of a varying $c$.

More seriously, however, all tests of a varying $c$ face a second
hurdle: the effects predicted are invariably either well beyond
the reach of current technology, or at best on the threshold.

In what follows we describe a number of observations which would either
provide positive detection of VSL effects, or imply constraints
upon the parameters of the theory. We also stick the neck out,
venturing a number of predictions of the theory.

\subsection{Changing-$\alpha$ and varying-$c$}\label{webbres}
\begin{figure}\label{fig4}
\begin{center}
\psfig{file=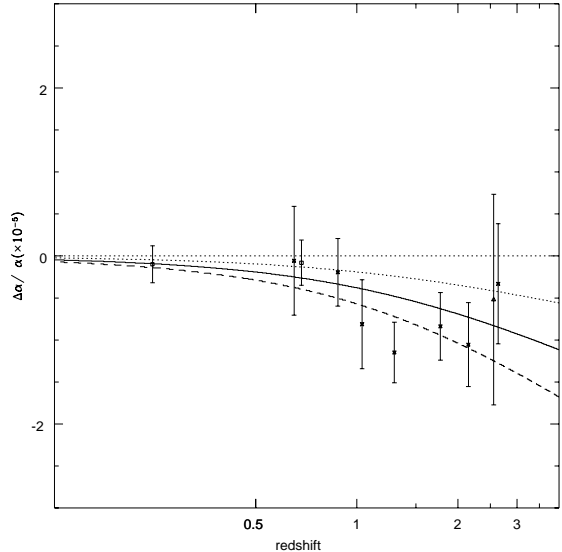,width=8cm} \caption{The data points are the
QSO results for the changing $\alpha$. The solid line depict
theoretical prediction in several varying-$\alpha$ models.}
\end{center}
\end{figure}

Perhaps the most extraordinary observation with relevance for VSL
is the work of Webb et al~\cite{webb}  and Murphy et
al~\cite{murphy} (hereafter referred to as the ``Webb results''
for historical reasons). These authors have reported evidence for
a redshift dependence in the fine structure constant
$\alpha=e^2/(\hbar c)$. This evidence was provided by a new
observational many-multiplet technique,  which exploits the extra
sensitivity gained by studying relativistic transitions to
different ground states, using absorption lines in quasar (QSO)
spectra at medium redshift. The trend of these results is that the
value of $\alpha $ was {\it lower} in the past, with $\Delta
\alpha /\alpha =(-0.72\pm 0.18)\times 10^{-5}$ for $z\approx
0.5-3.5$. These results are displayed in Fig.~\ref{fig4}. For
non-``fundamentalists'' (see Section~\ref{loophole} for a
definition), the obvious question is: If $\alpha$ is varying, what
else must be varying: $e$, $c$, $\hbar$, or a combination thereof?
Could such a matter be resolved by experiment?

As explained in Section~\ref{mean}, such a matter can only be
settled once a concrete theory is proposed, with a dynamics
capable of selecting a preferred system of units. This dynamics
may then be subjected to further experimental scrutiny.
Unfortunately the actual $\alpha(z)$ observations are not capable
of distinguishing between these theories. The dynamical equations
for variations in $\alpha$ invariably take the form \be
\label{dynalpha}\Box {\delta \alpha\over \alpha}=F(\rho) \ee In
varying $e$ theories \cite{bsbm,bek2,olive} one may have for
instance: \be F(\rho)=(2/\omega){\cal L}_{em}=
(2/\omega)\zeta_m\rho_m\ee 
where $\zeta$ is the ratio between $E^2-B^2$ and $E^2+B^2$.
In covariant VSL theories
\cite{covvsl}) one has \be F(\rho)=-(bq/\omega){\cal L}_M\ee and
in  VSL theories with hard breaking of Lorentz invariance one may
have \cite{vslsn}\be F(\rho)=-4\pi G \omega p(\rho).\ee In all
these cases the wave (Box) operator in a homogeneous Universe is
\be\Box \psi=-\ddot\psi-3{\dot a\over a}\dot \psi \ee where
$\psi\propto \delta \alpha/\alpha$, so the equations are formally
identical, of the form\be\label{dynalphacosmo} \ddot {\delta
\alpha\over \alpha}+3{\dot a\over a}\dot {\delta \alpha\over
\alpha} =-F(\rho) \ee and it is always possible to fit the Webb
results with appropriate parameters.

Notice, however, that although all these theories reveal the same
$\alpha(z)$, this depends upon the cosmological model, e.g. how
much cosmological constant is present. In all these theories
expansion acts as a friction force via the term in $\dot a/a$ in
Eqn~(\ref{dynalphacosmo}). Thus the current acceleration of the
universe cannot be ignored and has a clear imprint upon the
function $\alpha(z)$. Typically the onset of acceleration
suppresses variations in whatever ``varying'' constant.

This is of crucial importance when attending to other constraints
on varying alpha theories. Geonuclear tests, such as the Oklo
natural nuclear reactor \cite{sh,fujii} severely constrain
variations in $\alpha$ at low redshift. Thanks to the cosmological
constant varying alpha theories invariably
accommodate low redshift constraints, while explaining the Webb
results.

Given the non-discriminatory nature of $\alpha(z)$ we conclude
that to distinguish between varying $e$ and varying $c$ theories
one must look elsewhere. The status of the equivalence principle
in these theories turns out to be a good solution. The varying $e$
theories \cite{bsbm,bek2,olive} violate the weak equivalence
principle, whereas VSL theories do not\cite{mofwep,mbswep}. The
E\"otv\"os parameter
\begin{equation}
\eta \equiv {\frac{2|a_1-a_2|}{a_1+a_2}}
\end{equation}
is of the order $10^{-13}$ in varying $e$ theories, just an order
of magnitude below existing experimental bounds. The STEP
satellite could soon rule out the varying $e$ theories capable of
explaining the Webb et al results~\cite{step}.

For further varying alpha work in the context of brane world
cosmology see \cite{alphadou}; an analytical supergravity model
may be found in \cite{perry}. Also \cite{mofwep} is an example of
a varying alpha model based on a VSL theory in which $c(t)$ is
pre-given (see discussion in Section~\ref{hardvsl}).

\subsection{Supernovae results}
\begin{figure}
\begin{center}
\psfig{file=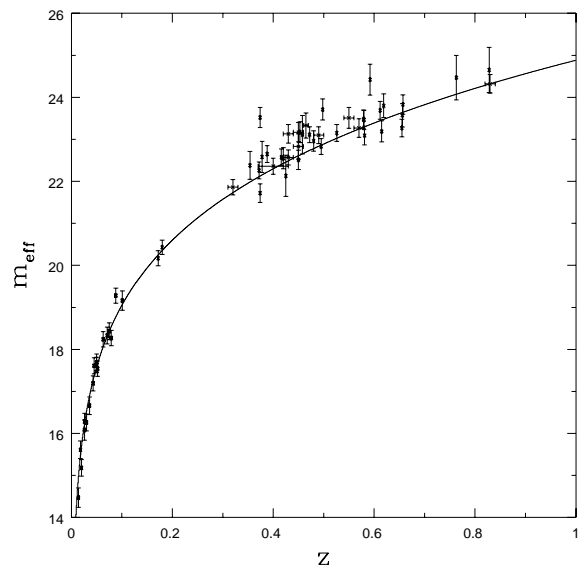,width=8cm} \caption{The Hubble diagram built
from Supernovae results (data points) suggests a Universe with
30\% normal matter and 70\% cosmological constant (plotted curve).
However any other form of repulsive gravity could be made to fit
the data.}\label{fig5}
\end{center}
\end{figure}

Recent astronomical observations of distant supernovae
light-curves have been realized by The Supernovae  Cosmology
Project and the High-z Supernova Search
\cite{super,super1,super2,super3}. These have extended the reach
of the Hubble diagram to high redshifts and provided evidence that
the expansion of the universe is accelerating (see
Fig.~\ref{fig5}).

A first question is how to interpret the Hubble diagram in VSL
theories.  Among other matters, the link between luminosity
distance and look-back time is obviously modified: with a higher
$c$ in the past objects with the same look-back time are further
away~\cite{bel,bel1,belinchon,vslsn,bimsn}.
If, however, one takes into account
the fine-structure results of Section~\ref{webbres}, one finds that any
corrections to the construction of the Hubble diagram must be very small
\cite{vslsn}, unless very contrived variations in $e$, $G$ and $\hbar$ are
introduced~\cite{bel2}. Therefore, even in VSL theories it looks as if
the universe {\it is} accelerating.

This {\it may} imply that there exists a significant positive
cosmological constant, $\Lambda $. If $\Lambda >0$, then cosmology
faces a very serious fine tuning problem, and this has motivated
extensive theoretical work. There is no theoretical motivation for
a value of $\Lambda $ of currently observable magnitude; a value
$10^{120}$ times smaller than the ``natural'' Planck scale of
density.

One possible explanation is VSL \cite{vslsn}. In such theories the
energy density in Lambda does not need to remain constant as in
the standard theory,  and thus does not require fine tuning.
Indeed it is possible to set up theories in which the presence of
Lambda drives changes in $c$, which in turn convert the vacuum
energy into ordinary matter. In such theories the supernovae
results can be explained without any need to fine-tune the initial
conditions, in fact with all parameters of the theory being of
order one~\cite{vslsn}.

In addition there is a strange, not often noted coincidence
between the redshifts at which the Universe starts accelerating
and those marking the onset of variations in $\alpha$. This
coincidence can be explained within the framework of these VSL
theories, for the reason cited above: acceleration acts as a
brake on any constant variation. Thus in \cite{vslsn} {\it both}
the Webb and supernovae results are fitted using the same set of
parameters.

We should however point out that the supernovae results can be
more modestly explained by quintessence, a replacement for the
cosmological constant not dissimilar from the inflaton or dilaton
fields~\cite{quint}.

%\subsection{The status of color-dependent $c$}
%\begin{figure}\label{fig2}
%\begin{center}
%\psfig{file=uhecr.ps,width=8cm} \caption{The flux of cosmic rays
%at high energies. The dashed line illustrates the GZK cut-off.}
%\end{center}
%\end{figure}

\subsection{Observations on the edge}
There are other relevant detections of variations in
supposed constants of nature, with obvious implications for VSL.
However, we should stress that the uncertainties here are larger
than in the results mentioned above. What follows merits interest
but also caution.

Variations in other ``fine structure constants'' (such as the weak
and strong $\alpha$) have not been
neglected~\cite{strongal,weakal}. In most VSL theories
(e.g.~\cite{covvsl}) all $\alpha$'s are bound to vary, since they
must remain proportional to each other, given their form
$\alpha_i=g^2/(\hbar c)$, where $i=em,W,S$ (electromagnetic, weak
and  strong). Variations in $\alpha_W$ from beta decay were
examined in \cite{beta}. The Oklo natural reactor implies that the
fermi constant $G_F$ could not have differed by more than 2\% a
few billion years ago. However, \cite{beta} claim to have found an
apparent discrepancy between the current value of $G_F$ and that
inferred from geochemical double-beta decay for 82Se; and also
between results for old and young minerals of 82Se and 130Te.
Although the Oklo constraints are themselves very model-dependent,
these results don't seem to fit well with each other or with any
varying constant theory. They depend on a large number of
geological assumptions.

A further claim for variation in a ``constant'' of nature was
obtained from vibrational and rotational atomic lines originating
in faraway systems. Naively one might think that such lines ``measure''
$\hbar$, but closer inspection reveals that they 
are sensitive to the dimensionless
parameter $\mu=m_p/m_e$, the ratio between the proton and
electron mass. A 3-sigma detection: \be {\Delta\mu\over
\mu}=(5.02\pm 1.82)\times 10^{-5} \ee at a redshift of order 3 has
been reported in \cite{mu}. Whether or not a 3-sigma result is to
be declared a detection is of course debatable, but such a
detection would have a strong bearing upon VSL theories.

Perhaps unsurprisingly, VSL theories have been proposed which impinge
upon the issue of the dark matter in the
universe~\cite{drum1,bass,bass1,laura}. For instance, the bimetric
theory presented in \cite{drum1} contains a length scale of
galactic size. For distances shorter than this scale, the theory
reduces to General Relativity, but beyond this scale gravity
becomes much stronger. The theory can explain the observed
galactic rotation curves, doing away with the need for galactic
dark matter. Likewise, the theory provides an explanation for the
observed value of Hubble's constant in relation to observed
matter, without dark matter or energy.

In an orthogonal approach VSL may be used to produce dark matter candidates,
such as in the theory proposed in \cite{laura} (as already
mentioned in Section~\ref{colvsl}). Here double-branched deformed
dispersion relations allow for high-momentum, low-energy particles
to be left over from the Big Bang. However these double-valued dispersion
relations appear to be forbidden by the criteria for non-linear
Lorentz invariance spelled out in Section~\ref{physdsr} (see also
\cite{leejoao1}).

The Pioneer anomaly is another result causing
consternation~\cite{pion}. Could gravity go so horribly wrong at
distances smaller than 100 AU? The fact that most VSL (or
varying-$e$) theories produce a fifth force~\cite{mbswep} might
lead one to think that VSL could explain the Pioneer anomaly.
However, the predicted effect in VSL is invariably very small. The
problem is to come up with a theory which fits Solar system data,
while predicting significant novelties at $20-40$ AU. This rules
out any $1/r^2$ type of force, and a successful theory should
somehow incorporate the solar system scale (just as the theory
in \cite{drum1} incorporates the galactic scale). 
For the VSL fifth force~\cite{mbswep} this can only be contrived 
by including an appropriate potential.

\subsection{Constraints on space-time varying $c$}

For every VSL theory it is important to explore
the range of constraints arising from standard tests of relativity.
This includes effects upon planetary orbits (eg. the precession
of the perihelion of Mercury), upon light (eg. gravitational light
bending, or the radar echo time-delay). These effects were studied,
for instance, in the model described in Section~\ref{livsl},
\cite{covvsl,vslbh}. Tight constraints upon the parameters
of this theory were derived and then imposed upon further
work. This is a prototype of work which should, and in most cases
has been done for all VSL theories proposed to date.

In addition one may explore general ``theory-free'' constraints.
As explained above, one can construct VSL models which not
only solve the homogeneity problem, but also produce a scale-invariant
spectrum of Gaussian fluctuations. In this sense there is consistency
with WMAP observations. In all these models, as in inflation,
the fluctuations are produced in the very early universe. Late-time
variations in $c$ would, in addition, affect the way in which
these primordial fluctuations are processed into the actual CMB
ansisotropies. For instance, the ionization history of the universe
would be changed~\cite{dopeak,reion,dopeakwmap}, since a different
value of $\alpha$ would affect the binding energy of hydrogen and
the cross section for Thomson scattering. With simplifying assumptions
one may then show how the Doppler peak positions would be modified,
and thus constrain variations in $\alpha$ at $z\approx 1000$ to
a few percent. As pointed out in \cite{dopeakwmap}
the existence of an early reionization epoch (as suggested by
MAP) may lead to considerably tighter constraints very soon,
as the ``reionization bump'' may be useful for the purpose
of constraining (or detecting) variations in $\alpha$.
Sensitivity to variations in  $\alpha$ as small as 0.1 \%
should in principle be achieved.

Another set of constraints arises from possible
violations of charge conservation in VSL and other
varying $\alpha$ theories\cite{landau,vucetich}. These constraints
are very model dependent (e.g. the equations in \cite{landau}
are tied to one specific way of implementing electromagnetism
in the theories in Section~\ref{hardvsl}; cf. \cite{covvsl}).
Even if one takes their particular implementation, one finds that
many $c(t)$ functions bypass these constraints easily.  But they
are interesting nonetheless, in particular for their
implications for baryogenesis~\cite{landau}.
Similar limits may be derived from possible
seasonal variations in Solar neutrino experiments~\cite{torres}.
Correlating charge conservation
constraints  with  violations of the equivalence
principle has also been examined~\cite{landau1}.

The tests above refer to theories predicting space-time variations
in $c$. Theories where the speed of light is color dependent have
quite a different observational status, which has already been
discussed. It concerns threshold anomalies and gamma ray timing
experiments (see~\cite{gammatime} for an update on the latter). We
have already pointed out that the two types of theories may be
overlaid.

\subsection{Sticking the neck out}

Any physical theory should stick the neck out and make experimental
predictions, and in this respect  VSL  has not been shy.

A very promising area is cold atom clocks~\cite{sortais}. The
search for the perfect unit of time has led to the the quest for
very stable oscillatory systems, leading to a gain, every ten years, of
about one order of magnitude in timing accuracy.
Cold atom clocks may be used as laboratory ``table-top''
probes for varying $\alpha$, with a current sensitivity
of about $10^{-15}$ per year.  For all varying alpha theories based
on an equation of the form (\ref{dynalphacosmo}) one finds a similar
prediction for the current yearly variation in alpha, once the Webb
results are fitted. It is found (e.g. by following a numerical integration
as described in \cite{bsbm}) that at present:
\be
{\dot\alpha\over\alpha}\approx 2.98\times 10^{-16} h \quad {\rm year}^{-1}
\ee
with $H_0=100 h$~Km~sec$^{-1}$~Mpc$^{-1}$, $\Omega_\Lambda=0.71$
and $\Omega_m=0.29$. For $h=0.7$ this
gives a fractional variation in alpha of about $2\times 10^{-16}$ per year,
which should soon be within the reach of technology.
Such an observation would be
an incredible vindication of the Webb results. On the other
hand this effect would become a further annoyance for those
concerned with  the practicalities of defining
the unit of time.

Note that although this prediction
is more or less the same for all causal varying $\alpha$ theories,
it does depend on cosmological parameters like $\Omega_\Lambda$, $H$,
or $\Omega$. For instance the cosmological constant is essential is
suppressing variations in $\alpha$ nowadays, as already explained above.
Without Lambda the current rate of variation in $\alpha$ would be
of the order $10^{-14}$ per year.

Spatial variations of $\alpha $ are likely to be significant
\cite{mbswep} in any varying alpha theory. For any causal theory
the dynamical equation for $\alpha$ will take form
(\ref{dynalpha}), that is the left hand side will be a wave
operator. This operator leads to the general equation
(\ref{dynalphacosmo}) in the context of time variations in an
expanding universe. However, near a static configuration of source
masses, the wave operator reduces to the spatial Laplacian, so
that we obtain a general linearized equation of the form
\begin{equation}
\label{dynalphabh} \nabla^2  {\delta \alpha\over \alpha}=F(\rho)
\end{equation}
Relative variations in
$\alpha $ near a star are therefore proportional to the local gravitational
potential. The exact relation between the change in $\alpha $ with
redshift and in space (near massive objects) is model dependent,
but eq.(\ref{dynalphabh}) provides the framework for predictions.
For instance, we have
\begin{equation}
{\frac{\delta \alpha }\alpha }=-{\frac{\zeta _s}\omega }{\frac{M_s}{\pi r}}%
\approx 2\times 10^{-4}{\frac{\zeta _s}{\zeta _m}}{\frac{M_s}{\pi
r}} \label{alphar}
\end{equation}
for a typical varying $e$ theory, but
\begin{equation}
{\frac{\delta \alpha }\alpha }=-{\frac{bq}\omega }{\frac{M_s}{4\pi r}}%
\approx 2\times 10^{-4}{\frac{M_s}{\pi r}}\;.  \label{alphar1}
\end{equation}
for the VSL theory described in Section~\ref{livsl}. Here $M_s$ is
the mass of the compact object, $r$ is its radius, and $\zeta$
is the ratio between $E^2-B^2$ and $E^2+B^2$. When $\zeta _m$ (for
the dark matter) and $\zeta _s$ (for, say, a star) have different
signs, for a cosmologically {\it increasing}
$\alpha $, varying $e$ theories predict that $%
\alpha $ should {\it decrease} on approach to a massive object.
And indeed one must have $\zeta _m<0$ in order to fit the Webb
results. In VSL, on the contrary, $\alpha $ {\it increases}
near compact objects (with decreasing $c$ if $q<0$, and increasing $c$ if $%
q>0$). In VSL theories, near a black hole $\alpha $ could become
much larger than 1, so that electromagnetism would become
non-perturbative with dramatic consequences for particle physics near
black holes. In varying-$e$ theories precisely the opposite
happens: electromagnetism switches off.

These effects are in principle observable using similar
spectroscopic techniques to those of Webb, but applied to lines
formed on the surface of very massive objects near us (in the
sense of $z\ll 1$). For that, we
need an object with a radius sufficiently close to its
Schwarzchild radius, such as an AGN, a pulsar or a
white dwarf, for the effect to be non-negligible. Furthermore we
need the ``chemistry'' of such an object to be sufficiently
simple, so that line blending does not become problematic.

Generally (i.e. for any matter configurations) the larger the
gravitational potential differences, the stronger the effect.
Indeed, for static configurations, both $\Delta\alpha/\alpha$ and
the gravitational potential satisfy Poisson equations, with source
terms related by a multiplicative constant. Hence the local value
of $\alpha$ should map the gravitational potential, and one would
need to have big variations in the gravitational potential to
observe corresponding spatial variations in $\alpha$.

Another area currently being studied are forces of inertia in
frame-dependent VSL theories (a matter briefly discussed in
Section~\ref{preframe}). High precision gyroscopes could soon put
these theories to the test.

\section{Upcoming attractions}

As the last section should have made clear, the future of VSL is
in the hands of observers. In such circumstances  theorists are
bound to continue with their musings. Some of these may lead to further
predictions and should be taken very seriously; others carry little more
than entertainment value. We close with an impressionistic
list of upcoming areas of  theoretical research.

One area in which not much work has been done is the relation
between VSL and quantum mechanics. Right from the start (e.g.~\cite{am})
it was obvious that a varying $c$ {\it implied} a varying $\hbar$.
This connection has  clear physical implications, for example particle
creation~\cite{covvsl,partcreat}. But there is more: it could be
that some of the strange properties of quantum mechanics -- such
as its non-locality~\cite{qmnonloc} -- are the result of a  form
of ``faster than light'' communication. Indeed, a VSL solution to the
problem of quantum entanglement has been proposed using the
bimetric VSL theory~\cite{qmnonloc1}. Such a theory removes the
tension between macroscopic classical and local gravity and
non-local quantum mechanics. On a different front,
it is also conceivable that the
non-locality of quantum mechanics could be used as a solution to
the homogeneity problem (and even more interestingly, the source
of cosmic structure). In this sense, quantum mechanics may already
be a VSL theory. However, a concrete theory based on this
idea is still missing.

Part of the reason for this intellectual black hole is the paucity
of work on VSL quantum cosmology. A possible excuse
arises from the exotic fact  that in some VSL theories
there is no quantum epoch!  In the model of \cite{am},
for instance, as we go back in time we find an {\it increasing}
$t/t_P$ (where we recall $t_P$ is the Planck time). Hence the
universe does not have a quantum origin, and one need not invoke
quantum cosmology to set its initial conditions.  Moreover, after
a sufficient number of Big Bang cycles, the value of $%
\Lambda $ in fundamental units will grow to be of order 1. Thus,
quantum cosmology is in the future, not the past.

Nevertheless this is not a generic feature of VSL models, and some
implications for quantum cosmology have been
considered~\cite{qcosm,coule}. Specifically, in \cite{qcosm} the
Wheeler-DeWitt equation under a varying $c$ was written down in
the mini-superspace approximation. The quantum potential was
obtained and the tunnelling probability examined in both the Vilenkin
and the Hartle-Hawking approaches. Thus a description for the quantum
birth of the Universe in VSL cosmologies was found. As J. W. Moffat had
already pointed out~\cite{moffat93} the ``problem of time'' in quantum
cosmology does not exist in VSL theories.

Another area where plenty of work is expected in the near future,
is the inclusion of gravity in the theories presented in
Section~\ref{qg}. These theories are motivated by the need to introduce
new invariant scales corresponding to the quanta of space and time.
They are phenomenological descriptions of quantum gravity. Thus it
is important that they can describe curved quantum space-time and not only
its gravity-free counterpart. As explained in Section~\ref{theoryqg} the issue
of gauging non-linear realizations of the Lorentz group depends
crucially on how to define position space in these
theories~\cite{djj}. The work in~\cite{grav} is based on the
simplest construction, and indeed is the only approach to bear fruit
so far.

However other approaches can and should be pursued. If one starts from
a non-linear realization of the Lorentz group in {\it real} space, then
the problem reduces to gauging non-linear representations, following
the steps of Kibble~\cite{kibble}, who showed that 
General Relativity is the gauge theory of the Poincar\'e
group. Yet another approach results
if position
space is recovered by means of a non-commutative
geometry~\cite{velnonc}. Then the issue is how to introduce
curvature into non-commutative structures. However
it has been suggested~\cite{eli} that such an
undertaking is impossible. Whatever the case it would be
good to have alternatives to~\cite{grav}, describing gravity
in the contect of non-linear relativity.

These are some wide avenues where the field is active,
but of course the future is unpredictable. Therefore we close with some
random ideas.

Perhaps VSL could ensue from theories with two time
dimensions, one of the form $c_1t_1$, the other $c_2t_2$. The
time compactification dynamics would then produce a varying $c$,
leading to another interesting realization of VSL.

It has been suggested that VSL may modify completely our
perception of space travel~\cite{covvsl,spacet}. Indeed VSL cosmic
strings~\cite{covvsl} could act as a source for spatial variations
in the speed of light, causing an increase in $c$ along the
string, within a possibly macroscopic radius (the string
core itself is always microscopic). Thus they create a tunnel
within which light travels at much higher speeds. This could make
it possible to travel extremely fast throughout the Universe
without the annoyances of time dilation (which could be made
negligible).

Regarding time (rather than space) travel, the implications of VSL
for the concept of time machines have never been explored.
However  causality is usually built into VSL theories.

It has
also been suggested that VSL has something to say about the
probability and quality of life in the universe~\cite{wheeler}.

%And much  more.

\section*{Acknowledgements}I would like to thank
Andy Albrecht, Stephon Alexander, Giovanni Amelino-Cameia, John
Barrow, Kim Baskerville, Bruce Bassett, Lluis Bel, Massimo Blasone, Mike
Duff, Ruth Durrer, Jurek Kowalski-Gilkman, Dagny
Kimberly, Jean-Luc Lehners, Jo\~ao Medeiros,
John Moffat, Paulo Pires-Pacheco, Levon Pogosian,
Lee Smolin, and Neil Turok for shaping my views on
the subject of this review. I'm also grateful to Anthony Valentini
for bringing the works of Poincar\'e~\cite{poincare} to my
attention -- they should be compulsory reading for every
physicist. Finally, warm thanks to the Perimeter Institute for
hiding me when I needed to run away to get any science done.

\end{document}